\documentclass[12pt,preprint]{aastex}

\begin{document}
\shorttitle{X-rays and triggered star formation} \shortauthors{Getman
et al.}

\slugcomment{Accepted for publication in the Astrophysical Journal
09/11/06}

\title{X-ray Study of Triggered Star Formation and Protostars in IC~1396N}

\author{Konstantin V.\ Getman, Eric D.\ Feigelson, Gordon Garmire,
Patrick Broos, Junfeng Wang}

\affil{Department of Astronomy \& Astrophysics, 525 Davey Laboratory,
Pennsylvania State University, University Park PA 16802}

\email{gkosta@astro.psu.edu}

\begin{abstract}

The IC~1396N cometary globule (CG) within the large nearby HII
region IC~1396 has been observed with the ACIS detector on board
the $Chandra$ X-ray Observatory. We detect 117 X-ray sources, of
which $\sim 50-60$ are likely members of the young open cluster
Trumpler~37 dispersed throughout the HII region, and 25 are
associated with young stars formed within the globule.  Infrared
photometry (2MASS and $Spitzer$) shows the X-ray population is
very young: 3 older Class III stars, 16 classical T Tauri stars, 6
protostars including a Class 0/I system. We infer a total T Tauri
population of $\sim 30$ stars in the globule, including the
undetected population, with a star formation efficiency of
$1-4\%$.

An elongated source spatial distribution with an age gradient
oriented towards the exciting star is discovered in the X-ray
population of IC~1396N, supporting similar findings in other
cometary globules. The geometric and age distribution is
consistent with the radiation driven implosion (RDI) model for
triggered star formation in CGs by HII region shocks.  The
inferred velocity of the shock front propagating into the globule
is $\sim 0.6$~km/s. The large number of X-ray-luminous protostars
in the globule suggests either an unusually high ratio of Class
I/0 vs. Class II/III stars, or a non-standard IMF favoring higher
mass stars by the triggering process.

The $Chandra$ source associated with the luminous Class 0/I
protostar IRAS~21391+5802 is one of the youngest stars ever
detected in the X-ray band.  We also establish for the first time
that the X-ray absorption in protostars arises from the local
infalling envelopes rather than ambient molecular cloud material.

\end{abstract}

\keywords{HII regions - ISM: globules - open clusters and
associations: individual (IC~1396) - stars: formation - stars:
pre-main sequence - X-rays: stars}

\section{Introduction \label{introduction_section}}

It has long been recognized that star formation in molecular
clouds can be triggered by ionization or shock fronts produced by
nearby massive stars \citep{Elmegreen77,Habing79,Elmegreen02}.
This leads to triggered star formation at the interface between
HII regions and molecular clouds and, on large scales, to the
sequential formation of star clusters in molecular cloud
complexes.  Two mechanisms for triggered star formation have been
discussed: the radiation driven implosion (RDI) model and the
`collect-and-collapse' model \citep[see review by][]{Elmegreen98}.
In the RDI model, photoevaporation of the cloud outer layers
induces a shock that compresses the cloud interior leading to
gravitational collapse
\citep{Reipurth83,Sugitani89,Lefloch94,Gorti02}. In the
`collect-and-collapse' model, the HII region compresses the cloud,
triggering the formation of self-gravitating cores on timescales
of $\sim 10^5$ yr \citep[see review by][]{Henney06}. Modern models
indicate that the time-dependent relationships between
dissociation, ionization, and shock fronts can be complex:
gravitational collapse can form stars in a thin dense layer
between the shock and ionization fronts propagating through the
cloud \citep{Hosokawa05,Hosokawa06,Zavagno06}.

Triggered star formation has been reported on the edges of large
HII regions in the Carina Nebula \citep{Smith00}, Orion clouds
\citep{Stanke02,Lee05}, M~16 \citep{Fukuda02}, M~17
\citep{Jiang02}, 30~Doradus \citep{Walborn02}, NGC~3603
\citep{Moffat02}, RCW~49 \citep{Whitney04}, W~5 \citep{Karr05},
the Gum Nebula \citep{Kim05}, and samples of more distant HII
regions \citep{Deharveng05}. Most of these are large star-forming
complexes where the geometry and conditions are difficult to
ascertain.

Bright-rimmed cometary globules (CGs) are simpler structures where
triggering processes may be active and thus offer an opportunity
for understanding better the mechanisms at work. These are small,
isolated clouds with dense cores surrounded by ionized rims facing
the exciting star and tails extending in the opposite direction
\citep[e.g.][]{Loren78,Sugitani91}. CGs likely originate as dense
clumps in the parental molecular clouds that have emerged after
the dispersion of lower density gas by ultraviolet radiation from
OB stars.

\begin{figure}
\centering
\includegraphics[angle=0.,width=6.0in]{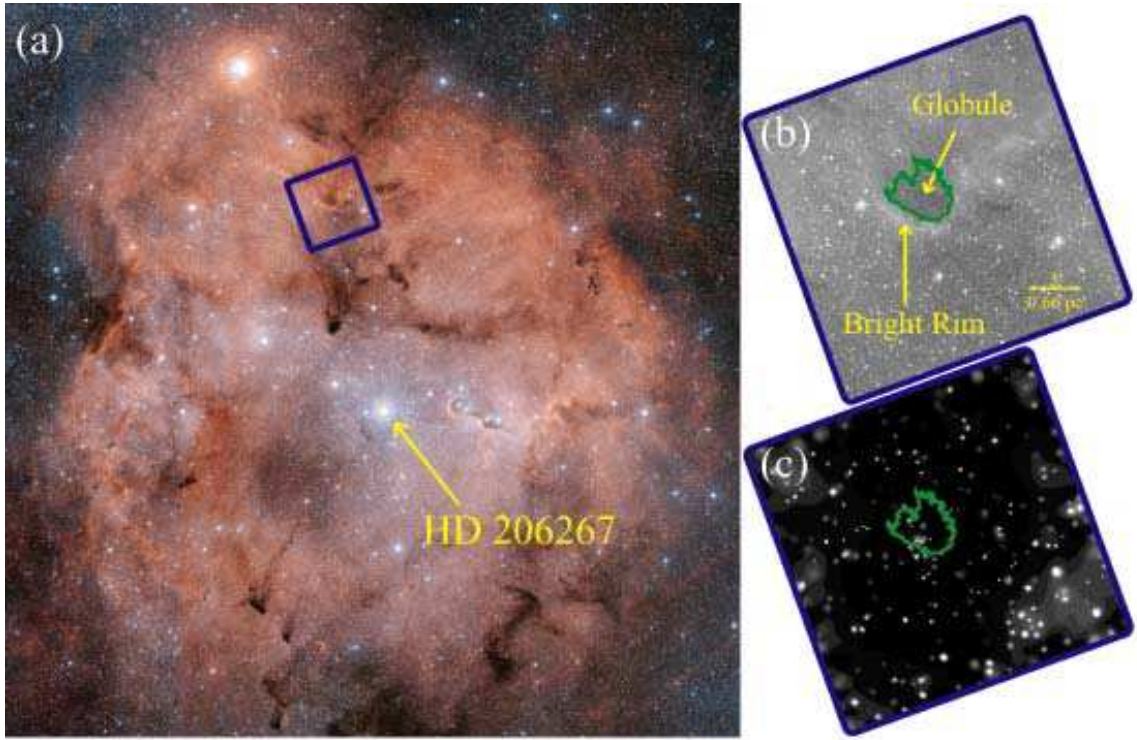}
\caption{(a) A $\sim 3\arcdeg \times 3\arcdeg$ composite (red $+$
blue) image of the emission nebula IC~1396 from the Digital Sky
Survey (DSS). The blue box outlines the $17\arcmin \times
17\arcmin$ $Chandra$ field centered on the bright-rimmed globule
IC-1396N. (b) A close-up $17\arcmin \times 17\arcmin$ DSS
grey-scale view of the globule. The green contour demarcates the
dark molecular head of the globule, tracing the ionization front
from HD~206267. (c) An adaptively smoothed $Chandra$ ACIS-I image
in the full $0.5-8.0$ keV band with the green contour superposed.
Smoothing has been performed at the 2.5$\sigma$ level, and gray
scales are logarithmic. Over 100 X-ray point sources are seen on
the image. \label{spat_distrib1_fig}}
\end{figure}

The large nearby HII region IC~1396 has a rich population of
bright-rimmed and cometary globules seen in silhouette against the
emission nebula (Figure \ref{spat_distrib1_fig}$a$).  Over a dozen
contain IRAS sources and are likely sites of star formation
\citep{Schwartz91}. Roughly 3$^\circ$ in diameter, IC~1396 is
excited by the O6.5f star HD~206267 in the Trumpler~37 cluster,
which lies at the center of the Cepheus OB2 Association
\citep{Simonson68,Walborn84}.

We study here globule IC~1396N \citep[labeled E in Figure 2 of
][]{Weikard96} located $\sim 11$~pc projected distance north of
HD~206267 with the bright rim tracing the ionization front facing
HD~206267 (Figure \ref{spat_distrib1_fig}$b$). Signs of ongoing
star formation in the globule include the luminous far-infrared
source IRAS~21391+5802, H$_2$O maser sources, molecular outflows,
HH flows, and clusters of near-infrared (NIR) embedded sources and
radio-mm protostars \citep{Sugitani89,Schwartz91,Tofani95,Slysh99,
Codella01,Nisini01,Beltran02,Reipurth03}. Mass estimates for the
globule range from 300 to 500~M$_{\odot}$ and the absorption
through its core is $A_V \sim 9-10$~mag
\citep{Wilking93,Serabyn93,Nisini01,Froebrich05}. We adopt a
distance of 750~pc \citep{Matthews79} for compatibility with most
recent studies, though we note that the $Hipparcos$ parallactic
distance measurement is $\sim 615$~pc \citep{deZeeuw99}.

Our study is unusual in that we use an X-ray telescope to find the
young stellar population around IC~1396N. X-ray surveys are
complementary to optical and infrared surveys because they trace
magnetic activity (mainly plasma heated in violent magnetic
reconnection flares) rather than photospheric or circumstellar
disk blackbody emission of low-mass pre-main sequence (PMS) stars
\citep[see reviews by][and references therein]{Feigelson99,
Feigelson06}. In regions like IC~1396, NIR surveys are overwhelmed
by unrelated field stars; many old background stars may penetrate
through the cloud and mimic young embedded objects (\S
\ref{nisini_cluster_subsection}). Foreground and background
Galactic stars have much less impact on X-ray studies, as magnetic
activity in PMS stars is elevated $10^1-10^4$ above main-sequence
levels \citep{Preibisch05a}. The main X-ray contaminants are
extragalactic objects, which are uncommon and can be identified
with some reliability \citep{Getman05b,Getman06}. X-ray surveys
penetrate deeply into obscuring material ($A_V > 100$~mag) and
thus are effective in detecting embedded objects
\citep{Getman05b,Grosso05}. X-ray observations also are not
hampered by two problems that affect NIR and optical studies of
young stellar populations in HII regions: they are not biased
towards stars with protoplanetary disks, and are not subject to
confusion from bright diffuse emission by heated gas and dust. The
{\em Chandra X-ray Observatory}, with its excellent
high-resolution mirrors, is often effective in resolving crowded
fields down to $\simeq 0.7$\arcsec\/ scales \citep{Getman05b}. The
central region of IC~1396 was previously studied with the lower
resolution $ROSAT$ observatory \citep{Schulz97} but the northern
part of the nebula with the IC~1396N globule was off the field.

The $Chandra$ observation of IC~1396N and source list are
described in \S \ref{observation_section}. Over 100 X-ray sources
are detected, from which we identify a cluster of 25 young objects
associated with the globule (\S \ref{globule_cluster_section}).
Based on location, absorption, and infrared photometry, we
distinguish embedded protostellar and more evolved T~Tauri
sub-populations (\S \ref{ir_prop_subsection}).  We compare the
X-ray cluster to the NIR observations of \citet{Nisini01}  in \S
\ref{nisini_cluster_subsection} and study X-ray properties of the
cluster in \S \ref{x_ray_prop_section}.  A particularly important
intermediate-mass Class 0/I protostar is presented in \S
\ref{srcsixtysix_section}. We end with discussion of how the rich
cluster of protostars in IC~1396N fits within the larger picture
of X-ray detections from protostars in other star forming regions,
and with discussion of the implications for understanding
triggered star formation in CGs (\S \ref{trigger_section}).

\section{The {\it Chandra} Observation and Source List
\label{observation_section}}

\subsection{Observation and Data Reduction \label{data_reduction_section}}

The X-ray observation of the globule IC~1396N and its vicinity was
obtained on $16.93-17.30$ October 2004 with the ACIS camera
\citep{Garmire03} on-board {\it Chandra} \citep{Weisskopf02}. We
consider here only results arising from the imaging array (ACIS-I)
of four abutted $1024 \times 1024$ pixel front-side illuminated
charge-coupled devices (CCDs) covering about $17\arcmin \times 17
\arcmin$ on the sky. The S2 and S3 detectors in the spectroscopic
array (ACIS-S) were also operational, but as the telescope point
spread function (PSF) is considerably degraded far off-axis, the
S2-S3 data are omitted from the analysis. The aim point of the
array was $21^{\rm{h}}40^{\rm{m}}42\fs4$,
$+58\arcdeg16\arcmin09\farcs7$ (J2000) or $(l,b) = (100.0, +4.2)$,
and the satellite roll angle (i.e., orientation of the CCD array
relative to the north-south direction) was $245\fdg9$. The total
net exposure time of our observation is 30 ks with no background
flaring due to solar activity or data losses.

Data reduction follows procedures similar to those described in
detail by \citet[][Appendix B]{Townsley03} and \citet{Getman05a}.
Briefly, the data are partially corrected for CCD charge transfer
inefficiency \citep{Townsley02}, cleaned with ``grade" (only ASCA
grades 0, 2, 3, 4, and 6 are accepted), ``status", and ``good-time
interval" filters, trimmed of background events outside of the
$0.5-8.0$ keV band and, cleaned of bad pixel columns with energies
of $< 0.7$ keV left by the standard processing. The slight PSF
broadening from the $Chandra$ X-ray Center's (CXC's) software
randomization of positions is removed. Based on several dozen
matches between bright $Chandra$ and 2MASS point sources, we
correct the absolute astrometry to the {\it Hipparcos} reference
frame.  These and later procedures were performed with $CIAO$
software package 3.2, $CALDB$ 3.0.0, $HEASOFT$ 5.3, CTI corrector
version 1.44, and the {\it ACIS Extract} package version 3.94. The
latter two tools were developed at Penn
State\footnote{Descriptions and codes for CTI correction and {\it
ACIS Extract} can be found at
\url{http://www.astro.psu.edu/users/townsley/cti/} and
\url{http://www.astro.psu.edu/xray/docs/TARA/ae\_users\_guide.html},
respectively.}.

\subsection{X-ray Point Source Catalog \label{srclist_section}}

\begin{figure}
\centering
\includegraphics[angle=0.,width=6.0in]{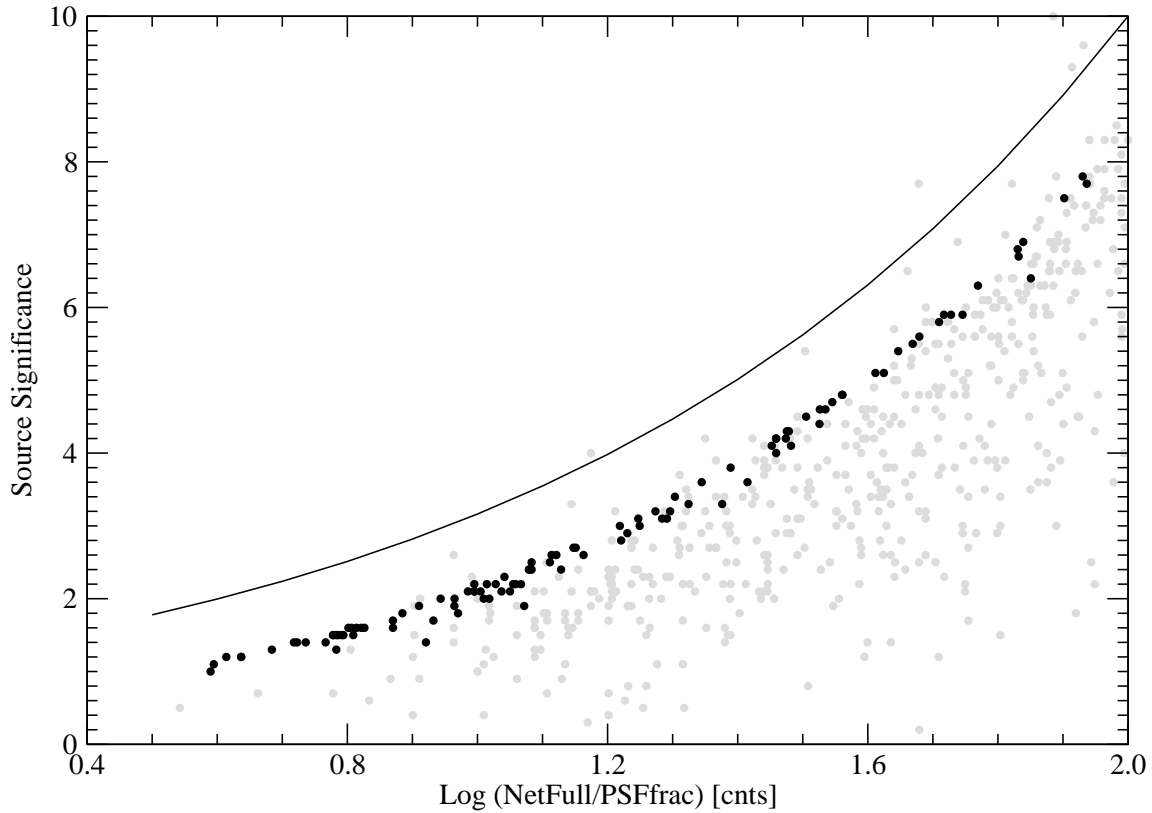}
\caption{Source significance versus source net counts. The solid
line shows the expected correlation between source significance
and net counts in a hypothetical case of no background. Black
circles are X-ray sources from the current $Chandra$ IC~1396
observation with a simple relatively flat random Poisson
background field. For comparison, grey circles show  Orion Nebula
sources from the COUP observation with a high and non-uniform
background field due to readout streaks and PSF wings from bright
sources. \label{sig_vs_counts_fig}}
\end{figure}

Figure \ref{spat_distrib1_fig}$c$ shows the resulting image of the
ACIS-I field after adaptive smoothing with the CIAO tool
$csmooth$. More than 100 point sources are easily discerned.
Source searching was performed with data images and exposure maps
constructed at three spatial resolutions (0.5, 1.0, and
$1.4\arcsec$ per pixel) using the $CIAO$ {\it wavdetect} tool. We
run {\it wavdetect} with a low threshold $P = 10^{-5}$ which is
highly sensitive but permits false detections at this point in the
analysis. This is followed by visual examination to locate other
candidate sources, mainly close doubles and sources near the
detection threshold. Using {\it ACIS Extract}, photons are
extracted within polygonal contours of $\sim 90\%$ encircled
energy using detailed position-dependent models of the PSF.
Background is measured locally in source-free regions. Due to a
very low, spatially invariant, ACIS-I background in the current
$Chandra$ observation of IC~1396 there is a one-to-one
correspondence between a source's significance and net counts
(black circles in Figure \ref{sig_vs_counts_fig}). For example,
this is not the case for a much more complicated COUP observation
\citep{Getman05a}, where the local source background is often
highly elevated due to crowded fields, readout trails, and PSF
wings from bright sources (grey circles in Figure
\ref{sig_vs_counts_fig}). The source significance, calculated by
{\it ACIS Extract}, is the ratio of the source's net
(background-subtracted) counts to the uncertainty on that
quantity\footnote{The uncertainty on the net counts is estimated
by calculating errors \citep[][eqs. 7 and 12]{Gehrels86} on total
observed and background counts and then propagating \citep[][eq.
1.31]{Lyons91} those errors to net counts. More detailed
information on the algorithm for calculating source significance
can be found in the section ``Algorithms'', subsection ``Broad
Band Photometry'', of the Users Guide for {\it ACIS Extract} at
\url{http://www.astro.psu.edu/xray/docs/TARA/ae\_users\_guide.html}.}:
$Signif = NetFull/\sigma(NetFull)$. Similar to the procedure of
\citet{Getman06}, the list of candidate sources is trimmed to omit
sources with fewer than $\sim 4$ estimated source
background-subtracted counts, $NetFull/PSFfrac \lesssim 4$.  In
this case the above criterion is equivalent to accepting sources
with a source significance of $Signif \ga 1$ (see Figure
\ref{sig_vs_counts_fig}). The resulting catalog of 117 sources is
listed in Table \ref{tbl_bsp}.

The {\it ACIS Extract} package estimates (often using $CIAO$
tools) a variety of source characteristics including celestial
position, off-axis angle, net and background counts within the
PSF-based extraction area, source significance assuming Poisson
statistics, effective exposure (corrected for telescope vignetting
and satellite dithering), median energy after background
subtraction\footnote{The derivation of median energy of
background-subtracted events is given in \citet[][\S
3.1]{Townsley06}.}, a variability indicator extracted from the
one-sided Kolmogorov-Smirnov statistics, and occasional anomalies
related to chip gap or field edge positions. These are all
reported in Table \ref{tbl_bsp}; see \citet{Getman05a} for a full
description of these quantities.

There are nine very weak sources with $3.9<NetFull/PSFfrac<5$
counts with low statistical significance.  Three of these (\#68,
72, and 82) will be discussed in \S \ref{globule_cluster_section}
as likely members of young stellar cluster in IC~1396N globule.
Four of these weakest sources (\#42, 57, 69, and 79) have very
hard spectra with median energy of $>3$~keV and without $2MASS$
counterparts; these are good candidates for extragalactic
contaminants.  The remaining two weak sources (\# 34 and 88) have
rather soft photon energies with $2MASS$ counterparts; these are
candidate members of the older cluster Trumpler~37 that extends
into the ACIS field of view. We thus believe most of these
faintest sources are real despite their marginal statistical
significance.

\subsection{Stellar and Nonstellar Sources \label{stellarID_section}}

Out of 117 X-ray sources, 66 are identified with 2MASS NIR
counterparts (last column in Table \ref{tbl_bsp}). Median
$Chandra-2MASS$ separations agree well with that derived from the
COUP project \citep{Getman05a}: they are better than $0.2\arcsec$
at the central part of the ACIS-I field, and $0.4\arcsec$ at
larger off-axis angles. Twenty nine of the 51 sources without NIR
detections have hard spectra with median energy $MedE > 3$ keV. Of
these, five are probably newly discovered protostars in the
globule (\S \ref{globule_cluster_section}).  The remaining 24 are
candidates for extragalactic contaminants (consistent with the
extragalactic prediction described below).

Figure \ref{spat_distrib2_fig}$a$ shows the spatial distribution
of the 117 X-ray sources on the ACIS-I field. There is a clear
spatial concentration of $\sim$ 25 X-ray sources (marked as
yellow, red, and blue) associated with the IC~1396N globule, which
we discuss below. But some of the 117 X-ray sources will be
unrelated to the IC~1396 star forming region.  To quantify this
problem, we use results from detailed simulations of Galactic
field and extragalactic background contamination populations
performed by \citet{Getman06} for the $Chandra$ observation of the
Cepheus B region, which had similar exposure time. The IC~1396 and
Cepheus B star forming regions have similar Galactic directions
($(l,b) = (100.0, +4.2)$ and $(l,b) = (110.2, +2.7)$ respectively)
with approximately the same distance from the Sun.  We thus expect
to have similar numbers of contaminating sources: $\sim 10$
Galactic foreground stars, $\sim 0$ Galactic background stars, and
$\sim 20$ extragalactic objects (considering all four CCD chips),
all uniformly distributed across the ACIS-I field. In total, we
estimate that $\sim 30$\% of the 117 sources are unrelated to
IC~1396 corresponding to a uniform density of one contaminating
source in every $\sim 10$ square arcminutes.

\section{X-ray Stellar Cluster in the Globule
\label{globule_cluster_section}}

\subsection{The X-ray sample \label{sample_subsection}}

We concentrate this study on sources which likely arise from star
formation in the globule, identified by their spatial clustering
within the globule and within the region around the bright rim
which traces the ionization front. About $50-60$ of the remaining
X-ray sources are probably members of the young open star cluster
Trumpler~37, part of the younger ($\sim 3$~Myr) subgroup of the
Cep~OB2 association \citep{Simonson68} surrounding the globule
(see Figure \ref{spat_distrib1_fig}$a$). They join the list of
previously unknown X-ray emitting low-mass PMS stars in the region
discovered with the $ROSAT$ satellite \citep{Schulz97}.

Figure \ref{spat_distrib2_fig}$b$ highlights the region around the
IC~1396N globule showing 30 X-ray sources on a 3.6 $\mu$m image
produced with the IRAC instrument on board the {\it Spitzer Space
Telescope}.  This region covers about 30.2 square arcminutes so
that, from the estimate in \S \ref{stellarID_section}, we expect
that a few of these 30 X-ray sources will be unrelated to the
cloud. Five sources (\# 46, 47, 50, 58, and 86) have properties
consistent with extragalactic contaminants \citep{Getman05b}: hard
X-ray spectra with $MedE >3$ keV, location around (rather than
within) the globule, and no 2MASS or $Spitzer$ counterparts. We
thus omit these sources and consider the remaining 25 as likely
stellar members of the globule.

\subsection{Infrared Photometry \label{IRphot_subsection}}

\begin{figure}
\centering
\includegraphics[angle=0.,width=3.8in]{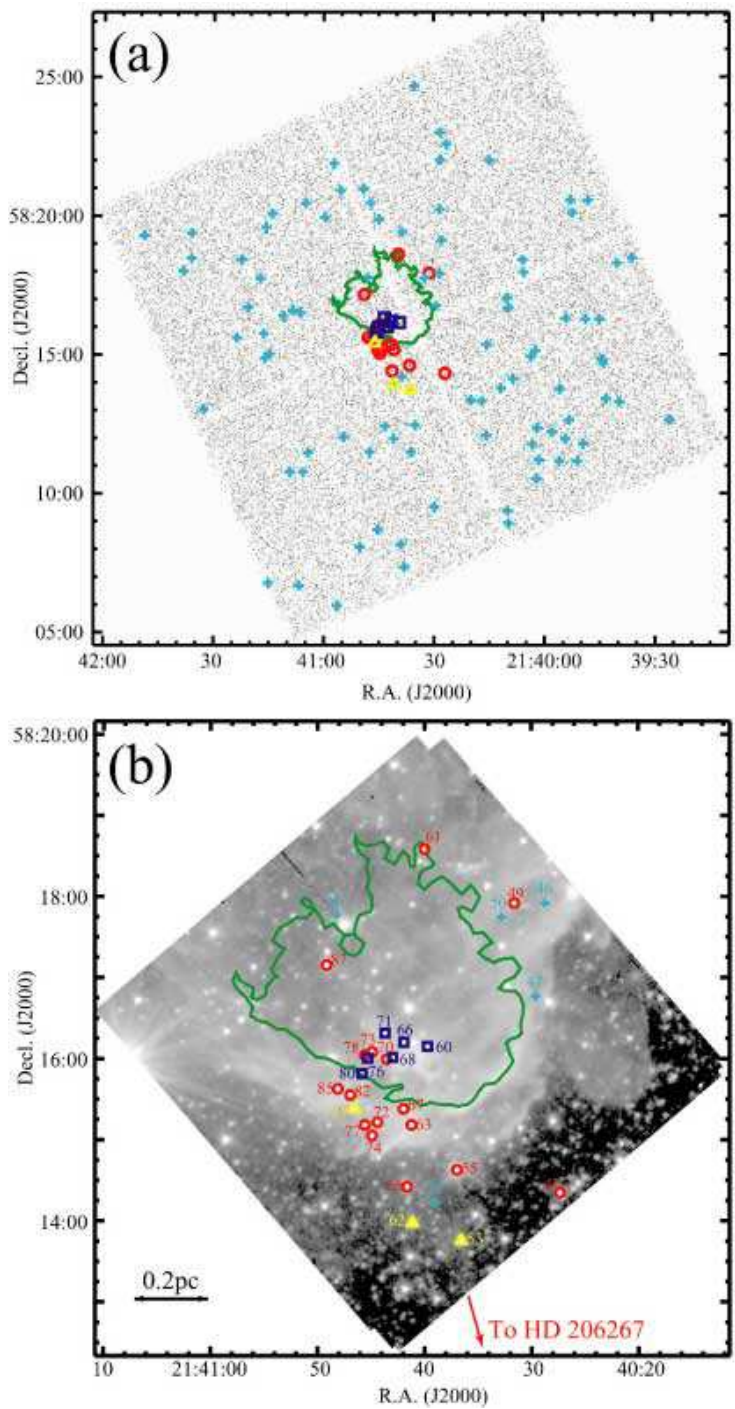}
\caption{(a) Spatial distribution of all 117 X-ray sources
superposed on a gray-scale low-resolution image of the $17\arcmin
\times 17\arcmin$ $Chandra$ field. (b) Close-up view of
$5.5\arcmin \times 5.5\arcmin$ around the IC~1396N globule from
the $Spitzer$ IRAC 3.6~$\mu$m image. The boundary of the globule
from Figure \ref{spat_distrib1_fig} is marked by the green contour
and the red arrow points towards HD~206267. In both panels, the 25
X-ray objects associated with the globule are marked by yellow
triangles (Class~III), red circles (Class~II) and blue squares
(Class~I/0). Older X-ray members of the dispersed PMS population
in the IC~1396 nebula and extragalactic and Galactic contaminants
are marked by cyan crosses. \label{spat_distrib2_fig}}
\end{figure}

The image in Figure \ref{spat_distrib2_fig}$b$ was one of several
retrieved from the $Spitzer$ archive data of IC~1396N after basic
calibration was applied.  MIR images of IC~1396N were obtained in
the 3.6, 4.5, 5.8, 8.0, 24 and 70 $\mu$m bands as part of the
observing program ``Star Formation in Bright Rimmed Clouds''
(Giovanni Fazio, PI). The {\it Spitzer Space Observatory},
Infrared Array Camera (IRAC), and Multiband Imaging Photometer for
Spitzer (MIPS) detectors are described by \citet{Werner04},
\citet{Fazio04}, and \citet{Rieke04}, respectively.

We extracted MIR photometry from the {\it Spitzer} data using two
different packages: the {\rm{\it apex.pl}} point source extraction
script (version 3.7) from the {\it Spitzer} MOPEX package; and the
{\it aper.pro} aperture photometry routine from the IDLPHOT
package\footnote{Descriptions and codes for MOPEX and IDLPHOT
packages can be found at
\url{http://ssc.spitzer.caltech.edu/postbcd/download-mopex.html}
and \url{http://idlastro.gsfc.nasa.gov/ftp/pro/idlphot/README},
respectively.}. In the first approach, source profile fitting is
performed using Point Response Functions (PRFs) which refines
source positions and automatically adjusts fluxes for crowded
sources. In the second approach, we use 5-pixel apertures for
single sources and 3-pixel apertures for weak crowded sources with
a flux correction factor of $1.08$ to account for photons outside
of the aperture. This small correction factor was derived by
comparing 3- and 5-pixel aperture photometry for a few bright
sources in the field.  Both methods give similar formal
statistical errors of typically $\sim 0.02$ mag, but for
individual sources their photometric results can differ up to
$\sim 0.1$ mag.

We report {\it apex.pl} photometry for the 25 X-ray sources
associated with the IC~1396N globule in Table \ref{tbl_nirmir}.
Zero magnitude flux densities of 280.9, 179.7, and 115.0 Jy have
been applied to IRAC fluxes in the 3.6, 4.5, and 5.8 $\mu$m bands,
respectively, to convert them to the magnitude system
\citep{Reach05}. We do not report photometry in 8.0 $\mu$m band as
sensitivity is reduced by bright extended emission attributed to
polycyclic aromatic hydrocarbons (PAHs).

Table \ref{tbl_nirmir} also gives $JHK$ photometry for 20 of the
25 X-ray sources with counterparts in the $2MASS$ catalog
(limiting magnitude $K_{s,lim} \sim 15.5$~mag) and the somewhat
deeper ($K_{lim} \sim 16.3$~mag) NIR observation of the globule by
\citet{Nisini01}. $Chandra-2MASS$ separations are generally better
than $0.2\arcsec$, indicating excellent X-ray astrometry similar
to that achieved in the Chandra Orion Ultradeep Project
\citep[COUP,][]{Getman05a}. X-ray sources with NIR counterparts
are shown as red and yellow circles in Figure
\ref{spat_distrib2_fig}$b$; these generally lie in the bright rim
of the globule.

\subsection{Infrared Properties of Cluster Members\label{ir_prop_subsection}}

\begin{figure}
\centering
\includegraphics[angle=0.,width=5.8in]{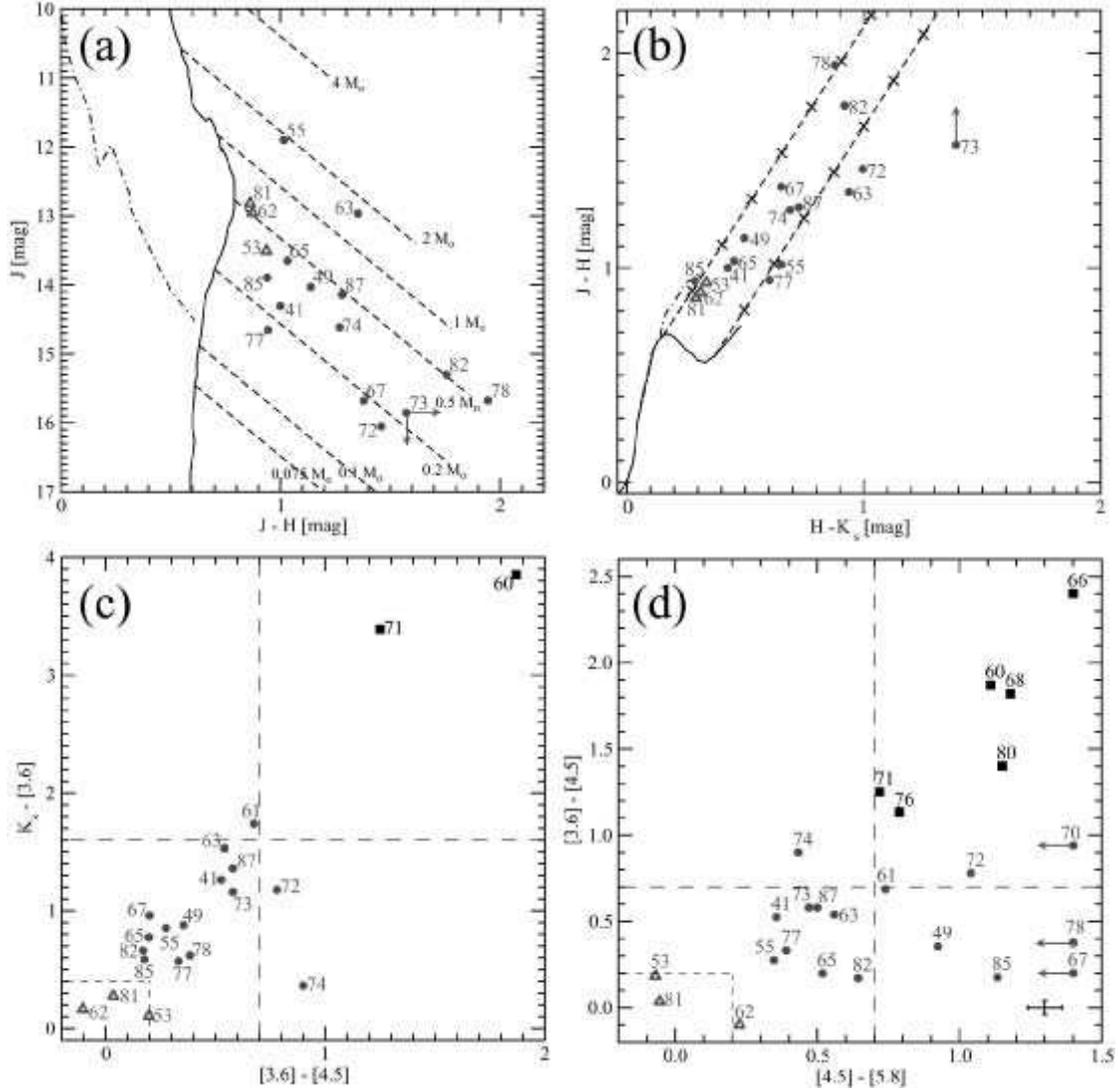}
\caption{Infrared properties of the 25 young X-ray objects
associated with the IC~1396N globule. Likely Class~III objects are
triangles, Class~II are circles, and Class~I/0 are squares. (a)
$2MASS$ NIR color-magnitude diagram. The unabsorbed ZAMS and
1\,Myr PMS isochrone are shown as dashed-dotted and solid lines,
respectively, from which $A_V \sim 10$ mag reddening vectors
(dashed lines) are shown for various star masses. The ZAMS is
obtained from Tables 7.6 and 15.7 of \citet{Cox00}, and the 1\,Myr
isochrone is from \citet{Baraffe98} for $0.02\, M_{\odot}
\leqslant M \leqslant 1.4\, M_{\odot}$ and \citet{Siess00} for
$1.4\, M_{\odot} \leqslant M \leqslant 7.0\, M_{\odot}$ assuming a
distance of 750 pc \citep{Matthews79}. Sources with unreliable or
absent $2MASS$ photometry are omitted. (b) $2MASS$ NIR color-color
diagram. The solid and dashed-dotted lines show loci of MS and
giant stars, respectively. The dashed lines are reddening vectors
originating at M0 V (left line) and M6.5 V (right line), and
marked at intervals of $A_V \sim 2$ mag. (c) NIR-MIR color-color
diagram based on $2MASS$ $K_s$ and two $Spitzer$ IRAC bands. (d)
MIR color-color diagram based on three $Spitzer$ IRAC bands.  In
panels (c) and (d), the dashed lines $K_s-[3.6] = 1.6$,
$[3.6]-[4.5] = 0.7$, and $[4.5]-[5.8] = 0.7$ discriminate
protostars from T-Tauri stars, and $K_s-[3.6] = 0.4$, $[3.6]-[4.5]
= 0.2$, and $[4.5]-[5.8] = 0.2$ --- Class~III from Class~II.
\label{nir_mir_fig}}
\end{figure}

Using the photometric results in Table \ref{tbl_nirmir}, we place
the globule X-ray sources on infrared color-magnitude and
color-color diagrams in Figure \ref{nir_mir_fig}.  Most of the
X-ray sources appear to be PMS stars with masses of $\sim
0.2-1$~M$_{\odot}$ subject to $A_V \sim 3-4$~mag absorption. A few
sources lying inside the head of the globule (e.g. \# 73, 78, and
82 in Figure \ref{spat_distrib2_fig}$b$) have high absorptions up
to 10 mag. In the $JHK$ color-color diagram (Figure
\ref{nir_mir_fig}$b$), a few sources (\# 63, 72, and possibly 55,
73, and 77) show the $K$-band excess indicator of a heated inner
dusty protoplanetary disk characteristic of accreting Class II PMS
stars \citep[e.g.][]{Kenyon95}.  Most have $JHK$ colors consistent
with the reddened photospheres characteristic of Class III PMS
stars, as is typically found in X-ray surveys of young stellar
clusters.

However, it is now recognized that $K$-band excess is a less
reliable indicator of disks and accretion than excesses in MIR
bands. $K$-band excess strongly depends on star-disk system
parameters such as the disk inclination, accretion rate, and inner
disk holes; it can be produced by other phenomena such as extended
emission in H~II regions and reflection nebulosity. Color indices
towards longer MIR wavelengths give more effective measures of
protoplanetary disks and better distinguish PMS phases: Class I
protostars, Class II accretion disk systems, and Class III systems
with weak or absent disks. Observations of Taurus-Auriga young
stars with reliable photometry show that the color criteria
$K_s-[3.6]>1.6$ and $[3.6]-[4.5]>0.7$ give excellent
discrimination between Class II and Class 0 or Class I systems
\citep{Hartmann05}.  Similarly, the criterion $[4.5]-[5.8]>0.7$
isolates Class 0 and Class I protostellar systems from Class II
and Class III T~Tauri systems.

$K_s-[3.6]$ vs. $[3.6]-[4.5]$ and $[3.6]-[4.5]$ vs. $[4.5]-[5.8]$
color-color diagrams are shown in Figure \ref{nir_mir_fig}$c-d$.
Based on the criteria of \citet{Hartmann05}, we classify 6 out of
25 X-ray sources in the IC~1396N cluster as Class 0 or Class I
protostars\footnote{Five out of six protostellar objects are
assigned to Class I, and one object (\#66), for which additional
(sub)millimeter data suggest a high ratio $L_{smm}/L_{bol}$ (\S
\ref{IRAS_subsection}) and a location within the transitional
Class 0/I locus on the evolutionary diagram for protostars (\S
\ref{xrays_protostars_section}), is assigned to Class 0/I.} which
we plot as squares in Figures \ref{spat_distrib2_fig} and
\ref{nir_mir_fig}. Sources \# 53, 62 and 81 with negligible
infrared excesses in all color-color diagrams can be best
classified as Class III PMS stars. These are plotted as triangles.
The remaining 16 X-ray sources are Class II `classical' T~Tauri
systems plotted as circles. Classifications based on MIR colors
are listed in column 15 of Table \ref{tbl_nirmir}.

\begin{figure}
\centering
\includegraphics[angle=0.,width=6.5in]{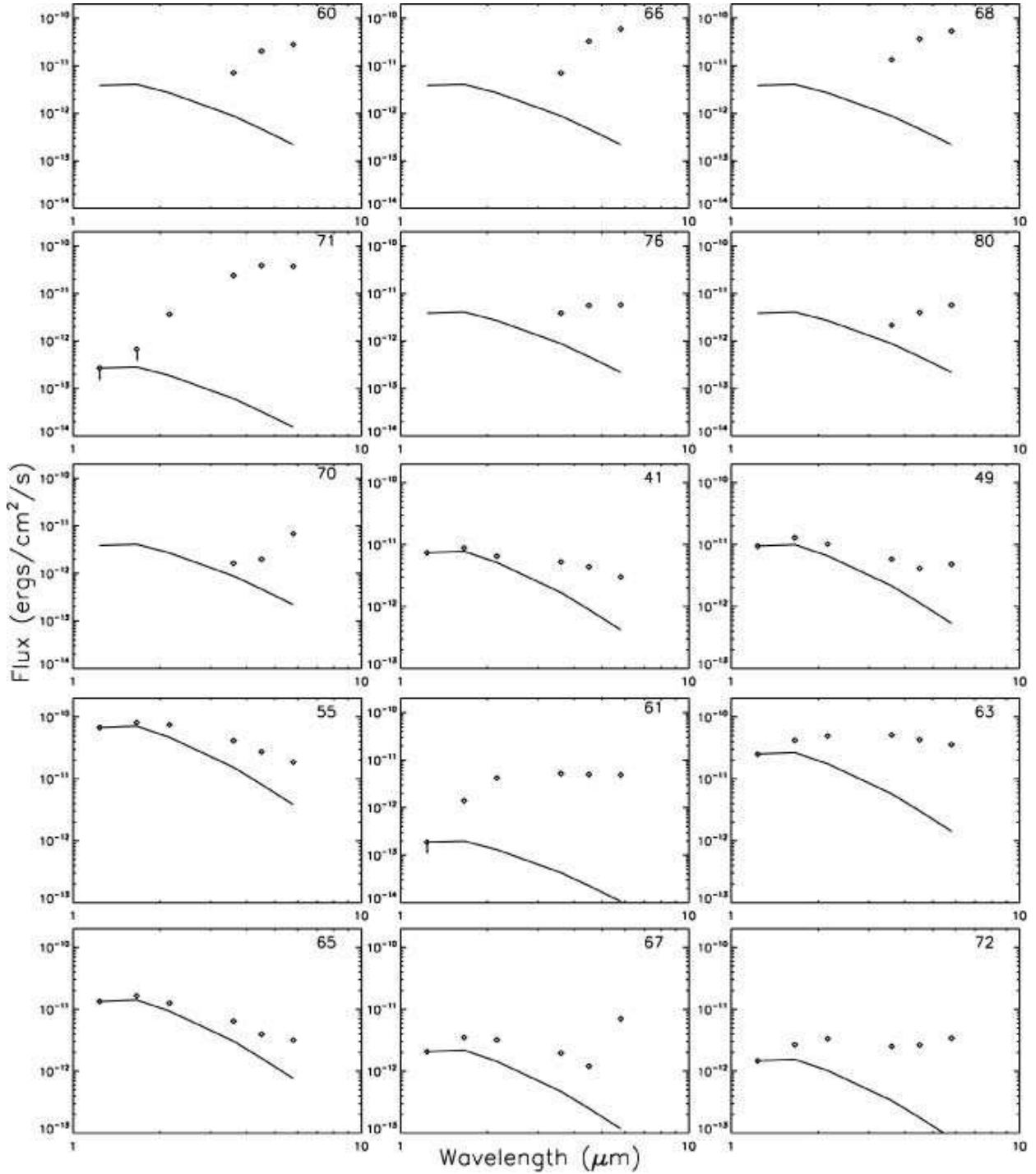}
\caption{Infrared SEDs of the 25 young X-ray objects associated
with the IC~1396N globule. $2MASS$ $JHK$ fluxes are combined with
$Spitzer$ fluxes in three shortest IRAC bands. The solid lines
reproduce (with arbitrary scaling) the SED of the Class~III source
\#81 as a typical stellar photosphere emission. Panels follow the
evolutionary sequence Class I-II-III as described in text.
\label{ir_sed_fig}}
\end{figure}

To further elucidate their infrared properties, we present in
Figure \ref{ir_sed_fig} the infrared spectral energy distributions
(SEDs) for all 25 X-ray stars in the IC~1396N globule. Up to six
bands ($2MASS$ $JHK$ and IRAC 3.6, 4.5, and 5.8 $\mu$m) are shown.
The panels show the proposed evolutionary sequence with the 6
Class 0 or Class I objects plotted first and the 3 Class III
objects plotted last. The SEDs largely confirm the classification
derived above from color-magnitude diagrams and emphasize some
interesting disk features in a few objects. All protostellar
objects (\# 60, 66, 68, 71, 76 and 80) have positive slopes of
their SEDs with increasing flux towards longer wavelengths. Source
\# 70, classified as Class II but without a NIR counterpart, is
shown next as a likely borderline Class I/II star.  The Class II
stars have roughly flat spectral slopes.  Source \#85 shows a
small excess at 5.8 $\mu$m  suggesting a disk inner hole, and \#67
and 78 have kinked SEDs suggesting flaring disks. The three Class
III stars (\# 53, 62 and 81) are consistent with simple
photospheres.

Four out of the six protostars are not seen in the NIR bands with a
limiting sensitivity of $K \sim 16.3$ mag \citep{Nisini01}. Two these
protostars, \#66 and 68, are identified here with young embedded
radio/mm objects, BIMA~2 and BIMA~3 \citep{Beltran02}. BIMA~2 $=$
\#66 is discussed further in \S \ref{srcsixtysix_section}.

\subsection{Comparison with the Nisini et al. sample
\label{nisini_cluster_subsection}}

Using the NIR camera SWIRCAM on the 1.1-m telescope AZT-24,
\citet{Nisini01} performed NIR observations of the IC~1396N
globule with the sensitivity limits of $(J,H,K)=(18.4,17.6,16.3)$
mag. The twenty reddest NIR sources ($J-H > 1.5$ mag and/or $H-K >
1.5$ mag) were proposed members of the embedded in the globule
young cluster. Table \ref{tbl_nisini} gives their positions and
$K$-band magnitudes from \citet{Nisini01}, their $Chandra$ and
$2MASS$ counterparts when available, and $Spitzer$ IRAC photometry
obtained by us with procedures as described in \S
\ref{IRphot_subsection}. We omit photometry in the [5.8] band as
most have low S/N ratios in this band.

\begin{figure}
\centering
\includegraphics[angle=0.,width=5.2in]{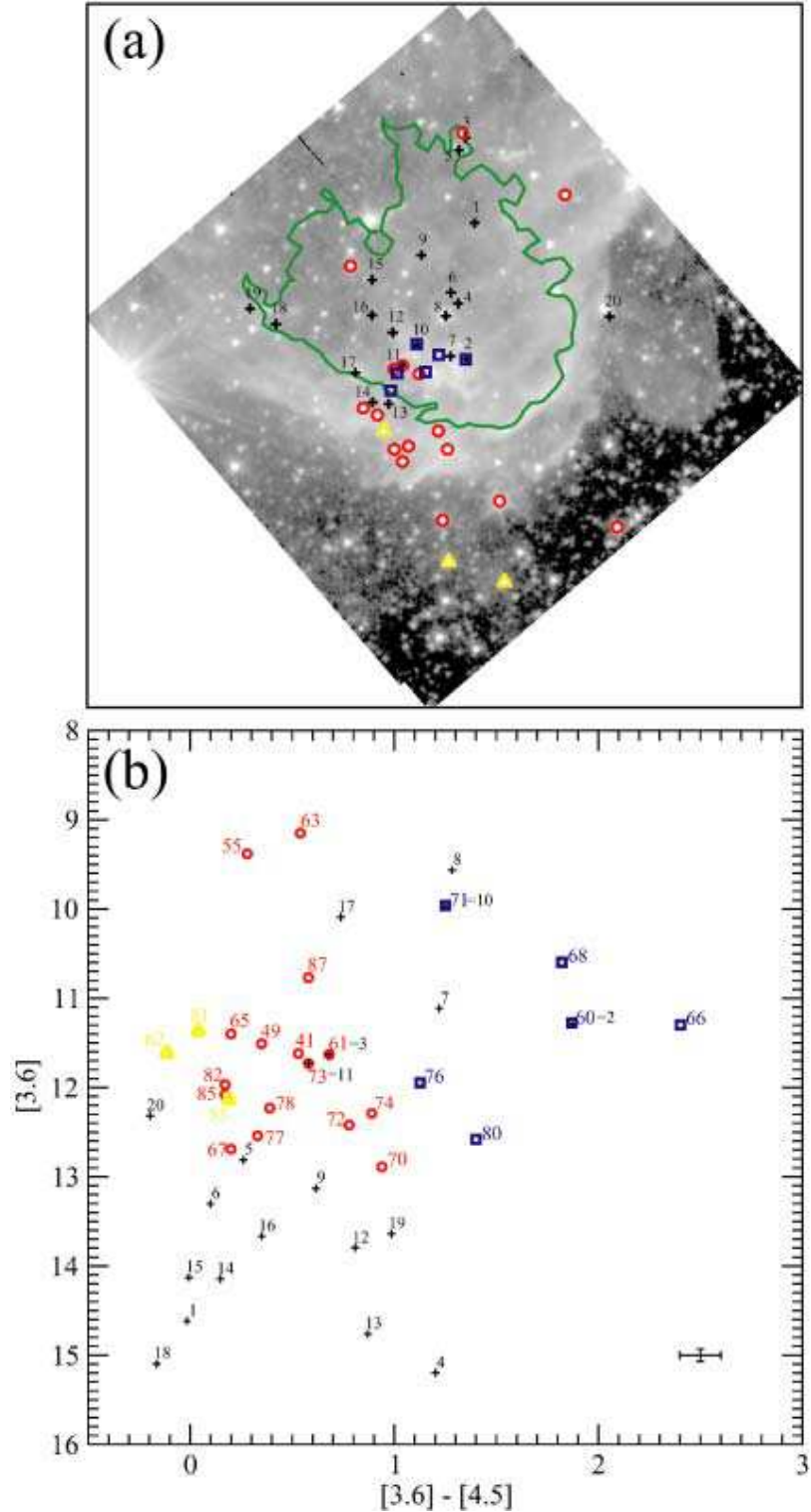}
\caption{(a) Source positions for the reddest NIR sources from
\citet{Nisini01} (black $+$ and labelled) and the X-ray cluster
overlayed on the $Spitzer$ IRAC image of the globule in 3.6$\mu$m.
(b) {\it Spitzer} color-magnitude diagram comparing colors of the
reddest NIR sources from \citet{Nisini01} (black $+$) with those
of the X-ray cluster. X-ray source symbols are the same as in
Figure~\ref{spat_distrib2_fig}. \label{nisini_fig}}
\end{figure}

Only four of these twenty sources from \citet{Nisini01} are
detected in our $Chandra$ observation. Figure \ref{nisini_fig}$a$
compares their spatial distributions. Most of the red NIR sources
are located northward along the same north-south line as the X-ray
objects, lying deep in the interior of the globule.  Figure
\ref{nisini_fig}$b$ shows that most of the NIR sources undetected
in X-rays are fainter than the $Chandra$ stars ($13 < [3.6] < 15$
mag compared to $[3.6] < 13$ mag). If they are members of the
globule, these Nisini et al.\ stars are less luminous and less
massive than the $Chandra$ sample. Indeed, combining a standard
Initial Mass Function with the $Chandra$ limit of $M \ga
0.4$~M$_\odot$ estimated in \S \ref{spectra_subsection}, a
population of undetected very low mass globule members is
expected. Alternatively, some of the Nisini et al.\ stars may be
reddened background stars seen through the cloud. Good candidates
for background stars are NIR stars (their \# 1, 5, 6, 14, 15, 18
and 20) with no intrinsic excess in either $K-[3.6]$ or
$[3.6]-[4.5]$ colors. \citet{Froebrich05} use the Besan\c{c}on
Galactic stellar population model to show that the background
stars in IC 1396 globules begin to exceed the $H-K = 0.8$ mag
level if the extinction of the cloud is higher than $A_V = 5$ mag.
A significant number of reddened background stars is expected for
globules with $A_V > 5$ mag; these are mostly M type giants for a
cloud with $A_V \sim 5$ mag, but other spectral types will
contaminate NIR samples of clouds with higher extinction.

Three of the \citet{Nisini01} NIR stars missing from our X-ray
image (their \# 7, 8 and 17) have color excesses consistent with
dusty disk systems and are located within the X-ray cluster. They
are also bright in the [3.6] band, with magnitudes far above the
sensitivity limit ($[3.6] \sim 13$ mag) of the $Chandra$
observation (Figure \ref{nisini_fig}$b$). These stars are probably
cluster members with X-ray emission below the sensitivity of our
short $Chandra$ exposure, or perhaps are viewed with their
circumstellar disks nearly edge-on in which case the X-ray
absorbing column may be increased \citep{Kastner05}.

We conclude that the brighter \citet{Nisini01} stars are very
likely globule members with dusty disks: four detected in X-rays
plus three bright in the [3.6] band, but undetected in X-rays.
Their fainter stars are either $M < 0.4$~M$_\odot$ cloud members
or background stars seen through the cloud. But perhaps of greater
importance is the failure of NIR methods to identify most of the
globule members found in the $Chandra$ image. The X-ray stars are
all easily seen on NIR images with $10<K<13.5$ mag, but their NIR
colors do not allow easy discrimination from the contaminating
population of Galactic field stars.

\section{X-ray Properties of the Cluster \label{x_ray_prop_section}}

Table \ref{tbl_xray} gives X-ray photometric and spectral
information on our 25 globule members. X-ray photometry is
obtained with $ACIS~Extract$ (\S \ref{srclist_section}) and
includes a number of useful quantities: number of source counts,
significance of a KS one-sample test for variability (${\rm
P}_{KS}$), and the background-corrected median X-ray photon energy
in the $(0.5-8.0)$ keV band ($MedE$).

\subsection{Flaring \label{flaring_subsection}}

Previous studies have shown that PMS stellar X-ray emission is
highly variable due to magnetic reconnection flaring
\citep{Feigelson06}. A flare with peak X-ray luminosity $\log L_x
\ga 30.0$ erg s$^{-1}$ occurs every few days in solar-mass Orion
PMS stars \citep{Wolk05}. Between flares, these stars spend 3/4 of
the time in a quasi-constant characteristic level that is likely
due to microflaring.  Flaring rates appear similar in lower mass
stars but with lower luminosities. Very powerful flares have also
been seen in Class I protostars \citep{Imanishi01}.

The present observation of IC~1396N has a duration of only $\sim
8$ hours, so only a small fraction of stars are expected to be
found in a flaring state.  Typical sources have only $10-30$
counts so that only high amplitude variations can be detected.
Table \ref{tbl_xray} shows that 3 of the 25 sources show
statistically significant (${\rm P}_{KS} < 0.005$) variability:
the protostar \#76 and the T~Tauri systems \#61 and 70. Tentative
evidence for X-ray variations are seen in another protostar (\#66)
and T~Tauri star (\#55).  Figure \ref{phot_arriv_fig} shows the
time-energy diagrams for photons detected from the six protostars.

\subsection{Absorption \label{absorption_subsection}}

X-ray median energy is a reliable indicator of absorbing column
density for sources with $MedE \ga 1.7$ keV, following the
empirical relationship $\log N_H~=~21.22~+~0.44 \times MedE$
cm$^{-2}$ \citep{Feigelson05}. All six X-ray protostars in
IC~1396N have $MedE \ga 4$ keV (Figure \ref{phot_arriv_fig})
corresponding to an absorbing column density of $N_H \ga
10^{23.0}$~cm$^{-2}$ (or $A_V \ga 50$ mag assuming a standard
gas-to-dust ratio of \citet{Ryter96}).

Due to the newly available $Spitzer$ photometry, we demonstrate
for the first time a relationship between X-ray $MedE$ and MIR
colors for obscured PMS systems.  Figure \ref{c_vs_mede_fig} shows
a clear correlation between the MIR color $[3.6]-[4.5]$ and the
X-ray median energy for all hard X-ray sources with $MedE > 3.0$
keV. A similar relationship is found using the [4.5]-[5.8] color
index. We show typical errors for the $Spitzer$ photometry and for
the $Chandra$ median energy.  The latter was derived from {\it
MARX} simulations for 10 count X-ray sources with a broad range of
spectral parameters \citep{Getman06}.

X-ray absorption at the energies of interest arises mainly from
inner shell photoelectric ionization in metals (N, O, Ne, Mg, Si,
Ar, Fe, etc.). Atoms in any physical state --- atomic, molecular
or solid --- contribute to the X-ray absorption. For PMS systems,
most of this gas is molecular and lies either in the intervening
cloud and/or in the circumstellar envelope or disk. The MIR
reddening arises from micron-scale solid-state interstellar
grains.

The specific pattern seen in Figure \ref{c_vs_mede_fig} can thus
be interpreted as follows.  For most T~Tauri systems (circles),
the scatter in the $[3.6]-[4.5]$ color is mainly due to different
accretion rates; sources with higher accretion rates are redder in
$[3.6]-[4.5]$ due to an increase in emission from the inner disk
and the heated `wall' at the dust sublimation edge
\citep{Allen04}. There is no report that higher accretion leads to
increased absorption of PMS X-rays; the scatter in $MedE$ for
T~Tauri stars most likely arises from different amounts of
intervening cloud material. At absorptions $\log N_H \la
22$~cm$^{-2}$, large differences in $N_H$ give relatively small
changes in $MedE$ for the $Chandra$ ACIS detector
\citep{Feigelson05}.   As the maximum line-of-sight column through
the IC~1396N globule is $A_V \sim 10$ mag (\S
\ref{introduction_section}), random locations in the cloud should
produce random scatter up to $MedE \sim 2.5$ keV, as observed in
Figure \ref{c_vs_mede_fig}.

For the protostars (squares), the MIR colors are mostly determined
by the density of the local infalling envelope.  The $[3.6]-[4.5]$
color index increases for denser envelopes because the $\tau \sim
1$ optical depth surface moves to the cooler outer envelope
\citep{Allen04}.  The proportional increase in $MedE$ seen among
the protostars in Figures \ref{c_vs_mede_fig} thus arises from
absorption by molecular gas in the local envelope.  This may be
the clearest demonstration to date of X-ray absorption from
protostellar envelopes independent of absorption in the larger
molecular cloud.

\subsection{X-ray luminosities, completeness, and the globule stellar population
\label{spectra_subsection}}

Results from the {\it ACIS~Extract} spectral analysis are given in
Table \ref{tbl_xray}.  Our procedures for fitting X-ray spectra
and obtaining broad-band luminosities are based on those described
in \citet{Getman05a,Getman06}. For the three brightest T~Tauri
sources with greater than 45 net counts (\#61, 62 and 63) and for
all protostars, one-temperature APEC plasma emission models were
fitted using the {\it XSPEC} package assuming 0.3 times solar
elemental abundances. To explore the whole range of various
alternative models that successfully fit the data, we performed
spectral fits on grouped (with different minimum numbers of
grouped counts, starting from 1 using the $\chi^2$-statistic) as
well as on un-grouped (using the C-statistic) spectra. We then
performed a visual inspection of all spectral fits to choose the
"most appropriate" spectral fit, in which the plasma energy most
closely matches the typical energy of PMS stars seen in the much
deeper Orion Nebula data \citep{Getman05a}. These typical energies
are $kT \sim 1.6$ keV for unabsorbed stars with NIR counterparts,
$kT \sim 2.6$ keV for absorbed stars with NIR counterparts, and
$kT \sim 4.4$ keV for absorbed stars without NIR counterparts. The
resulting $N_H$ values lie within $\pm 0.3$ dex of absorptions
derived from dereddening sources to the 1 Myr PMS isochrone
($N_{H,IR}$) in the $J$ vs. $J-H$ color-magnitude diagram (Figure
\ref{nir_mir_fig}$a$).

For the weaker T~Tauri sources in the sample, we assume the
absorbing column is the value $N_{H,IR}$ derived from NIR
photometry, and we estimate their plasma energy from $N_H - MedE$
plane using Figure~8 in \citet{Feigelson05}. With the $N_H$ and
$kT$ parameters frozen, we then performed $XSPEC$ fits to derive
the spectral model normalization and to calculate broad-band X-ray
fluxes.

Formal statistical errors on X-ray luminosities have been
estimated through Monte Carlo simulations described in \S 4.3 of
\citet{Getman06}. The scientific reliability of these
luminosities, however, is usually lower than the statistical
uncertainty for several reasons. First, the $L_{t,c}$ will
systematically underestimate the true emission for obscured or
embedded populations due to the absence of soft band photons.
Second, scatter is expected due to intrinsic variability (\S
\ref{flaring_subsection}). Third, if the distance of 750 pc
adopted in this paper were revised to the 615 pc value derived in
\citet{deZeeuw99}, then the PMS stellar X-ray luminosities will
systematically be decreased by 0.2 dex. The luminosity in the hard
$(2-8)$ keV energy band corrected for absorption ($L_{h,c}$) is
most suitable for description and comparison of the intrinsic
properties of young stellar objects between T~Tauri and
protostellar samples.

We show in Figure \ref{lmag_vs_lhc_fig} a MIR version of the
well-known correlation between X-ray luminosity and bolometric
luminosities for PMS stars \citep[e.g.][]{Preibisch05}.  In the
absence of complete detections in the NIR bands, we use the MIR
[3.6] IRAC band as a proxy for stellar bolometric luminosity.
$Chandra$ and $L\arcmin$ values for Orion Nebula T~Tauri stars
\citep[grey circles,][]{Getman05a}, adjusted to the distance of
the IC~1396N, are shown for comparison. Note that much of the
[3.6] or $L\arcmin$ emission can arise from circumstellar disks,
particularly in Class~I systems (Figure \ref{ir_sed_fig}), and the
stellar value will be lower than the plotted values.

The IC~1396N stars generally follow the distribution seen in Orion
stars, although three protostars and the Class~I/II system \#70
appear overluminous in X-rays.  It is not clear whether this
implies systematically higher X-ray luminosities in the youngest
systems, or the presence of still-undiscovered Class~I systems
with X-ray luminosities below the sensitivity limit of our short
$Chandra$ exposure.  If the T~Tauri $L_x -$mass relation extends
into the protostellar regime, lower masses exist within the
IC~1396N globule and would be detected with a more sensitive
$Chandra$ exposure.

The highest luminosities of T-Tauri stars in IC~1396N are $\sim
10^{30}$ ergs~s$^{-1}$, while the sensitivity limit of the
$Chandra$ observation for the sample is $L_{h,c} \sim 10^{28.5}$
ergs~s$^{-1}$. Using the COUP $L_x-$Mass correlation
\citep{Preibisch05}, the implied mass corresponds roughly to the
stellar mass range of the X-ray T~Tauri stars seen in IC~1396N of
$\sim 0.2-1$~M$_{\odot}$, which is similar to that obtained from
NIR color-magnitude diagram in \S \ref{ir_prop_subsection}
assuming an age of $\sim 1$~Myr. Comparison with the Cep~B cloud
\citep{Getman06} is helpful here because its distance, absorption,
and $Chandra$ exposure are similar to that of IC~1396N.  Through a
detailed analysis of the X-ray Luminosity Function (XLF) and
Initial Mass Function (IMF), we showed that the X-ray unobscured
population in Cep OB3b is complete down to $0.4$~M$_{\odot}$
(assuming an age of 1~Myr). We expect the IC~1396N observation to
be similarly nearly complete for masses $M>0.4$~M$_\odot$.  It
should capture roughly half of members in the range
$0.2<M<0.4$~M$_\odot$ and few members with $M<0.2$~M$_\odot$.

Although, the total population in IC~1396N is too small to permit
a detailed XLF and IMF analysis, we can estimate the total number
of unobscured stars (T-Tauri population) in the globule by
comparing a ratio of numbers of stars in two mass ranges
$0.1<M<0.4$~M$_\odot$ and $M>0.4$~M$_\odot$ for IC~1396N to that
of the COUP Orion unobscured sample. COUP is complete down to
$0.1$~M$_\odot$ with the ratio of $\sim 400$ to $\sim 300$ stars
\citep[\S 7.2 in][]{Getman06}. This implies that IC~1396N contains
$\sim 10$ T-Tauri stars with $M>0.1$~M$_\odot$ which are
undetected in our $Chandra$ observation, in addition to the
detected 19 members.  The total stellar mass of the unobscured
population (not including brown dwarfs or close binary companions)
for the globule is around $\sim 15$~M$_\odot$.

\section{X-rays from the Remarkable Protostar IRAS~21391+5802
\label{srcsixtysix_section}}

\subsection{Previous studies of IRAS~21391+5802\label{IRAS_subsection}}

IRAS~21391+5802, the most luminous MIR source in the IC~13906N
globule, is an embedded protostar producing a bipolar molecular
outflow \citep{Sugitani89, Saraceno96a}. From JCMT photometry and
ISO satellite spectrometry, \citet{Saraceno96} derive a bolometric
luminosity of 235~L$_{\odot}$. \citet{Nisini01} obtain the
$L_{smm}/L_{bol}$ ratio of $4.6 \times 10^{-2}$, where $L_{smm}$
is the luminosity emitted longward of 350$\mu$m.  This suggests
the source may be classified as an extremely young Class 0
protostar, according to the criterion $L_{smm}/L_{bol} > 0.5\%$
recommended by \citet{Andre00}. The millimeter emission is
resolved into two sources, A and B, by the OVRO array
\citep{Codella01}.

The highest-resolution studies of the source have been made with
the BIMA millimeter array and the VLA at centimeter wavelengths by
\citet{Beltran02}, resolving the emission into three sources
separated by $\sim 15\arcsec$. BIMA~2 (= VLA~2 = OVRO~A) is
associated with the IRAS~21391+5802 MIR emission and drives the
strong molecular outflow. The circumstellar mass from BIMA~2 is
$\sim 5$~M$_\odot$ in a very extended envelope and bolometric
luminosity is 150~L$_{\odot}$, while the other two components have
envelopes with $< 0.1$~M$_\odot$.  Taking all available
information into account, Beltr\'an et al.\ classify BIMA~2 as an
embedded intermediate-mass protostar likely to emerge as a main
sequence A- or B-type star. The source fits the correlation of
normalized outflow momentum and envelope mass found for low-mass
Class 0 objects \citep{Bontemps96}. BIMA~1 (which powers a small
bipolar outflow) and BIMA~3 are lower mass, and probably more
evolved, objects.

\subsection{$Chandra$ and $Spitzer$ properties \label{Chandra66_subsection}}

\begin{figure}
\centering
\includegraphics[angle=0.,width=6.5in]{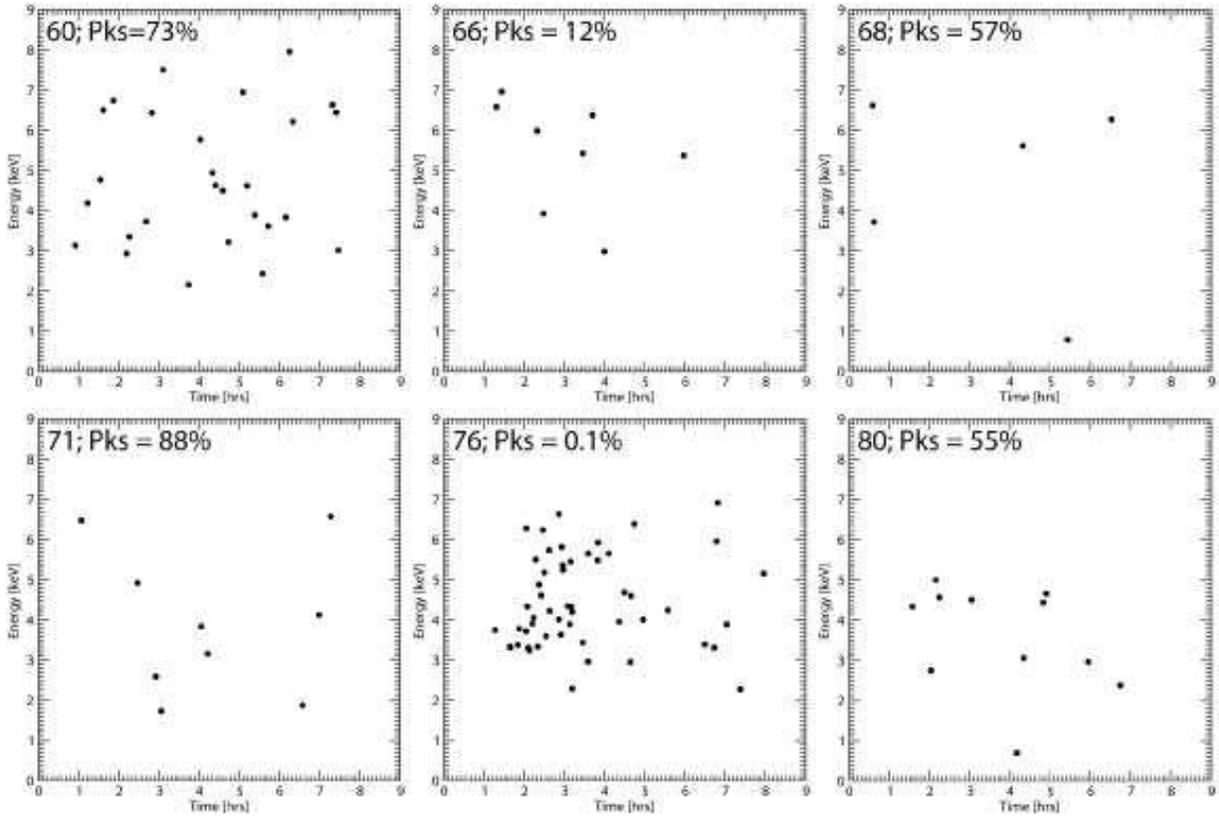}
\caption{X-ray photon-energy diagrams for the six protostars in
the globule. The quantity ${\rm P}_{KS}$ is the Kolmogorov-Smirnov
probability of accepting the null hypothesis of a constant source.
\label{phot_arriv_fig}}
\end{figure}

\begin{figure}
\centering
\includegraphics[angle=0.,width=6.5in]{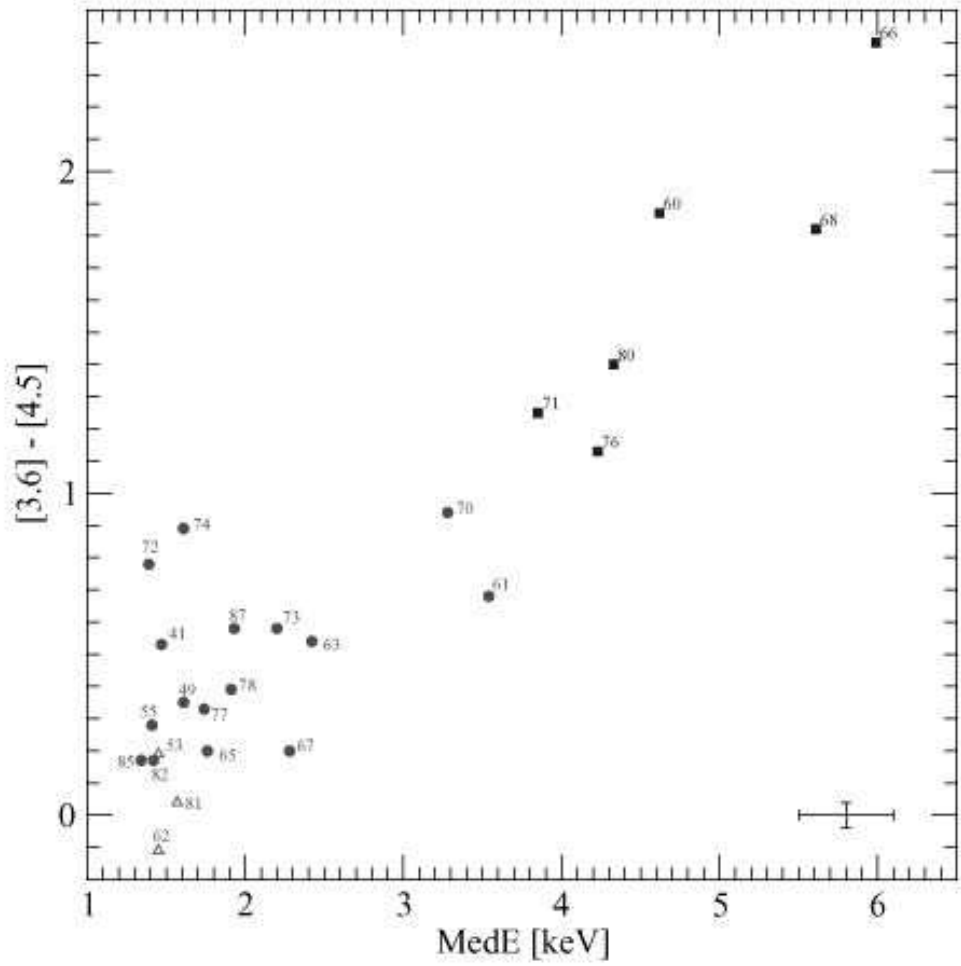}
\caption{{\it Spitzer} [3.6]-[4.5] MIR color versus {\it Chandra}
X-ray source median energy.  Symbols are the same as in
Figure~\ref{nir_mir_fig}. Typical error bars for an X-ray source
with 10 net counts are shown at the bottom right corner.
\label{c_vs_mede_fig}}
\end{figure}

\begin{figure}
\centering
\includegraphics[angle=0.,width=6.5in]{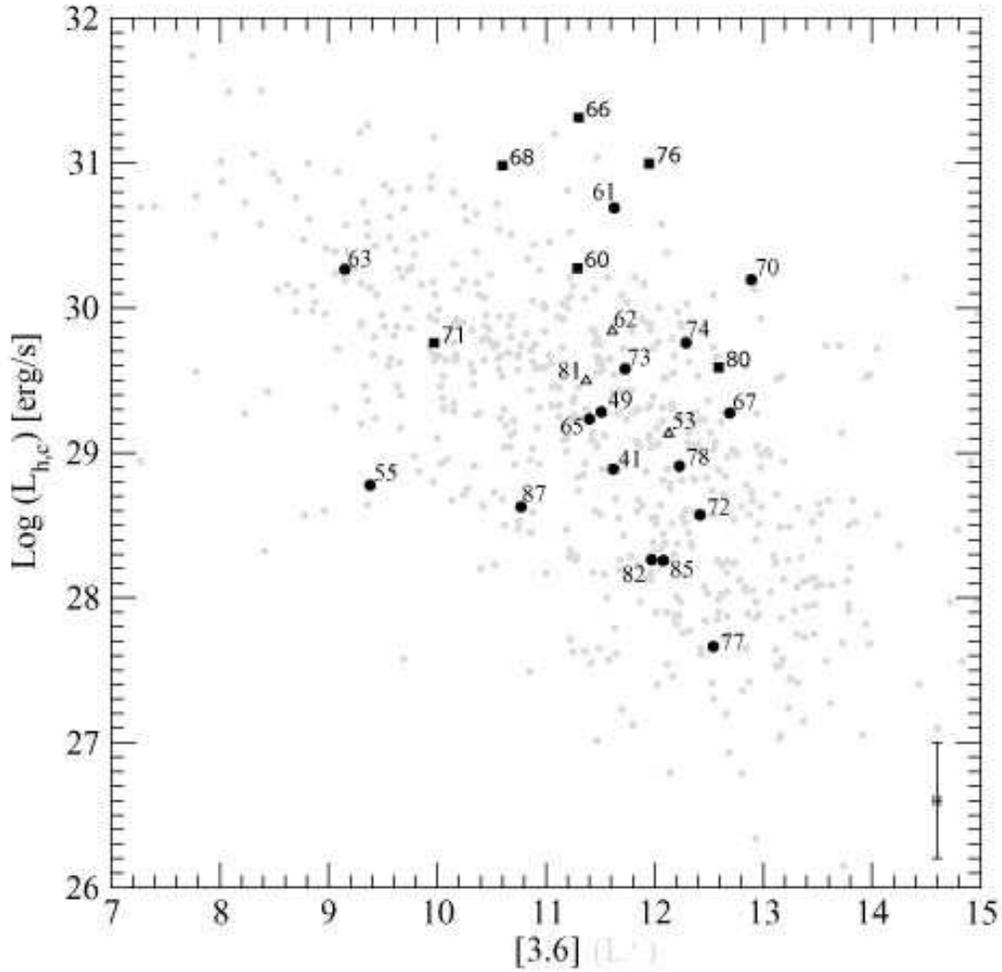}
\caption{$\log L_{h,c}$ ($(2-8)$~keV luminosities corrected for
absorption) plotted against the observed [3.6]~band magnitudes for
IC~1396N X-ray stars.  Symbols are the same as in
Figure~\ref{nir_mir_fig}. Grey circles show for comparison Orion
Nebula sample from the COUP project \citep{Getman05a} with
infrared magnitudes in $L^{\prime}$-band. \label{lmag_vs_lhc_fig}}
\end{figure}

\begin{figure}
\centering
\includegraphics[angle=0.,width=2.8in]{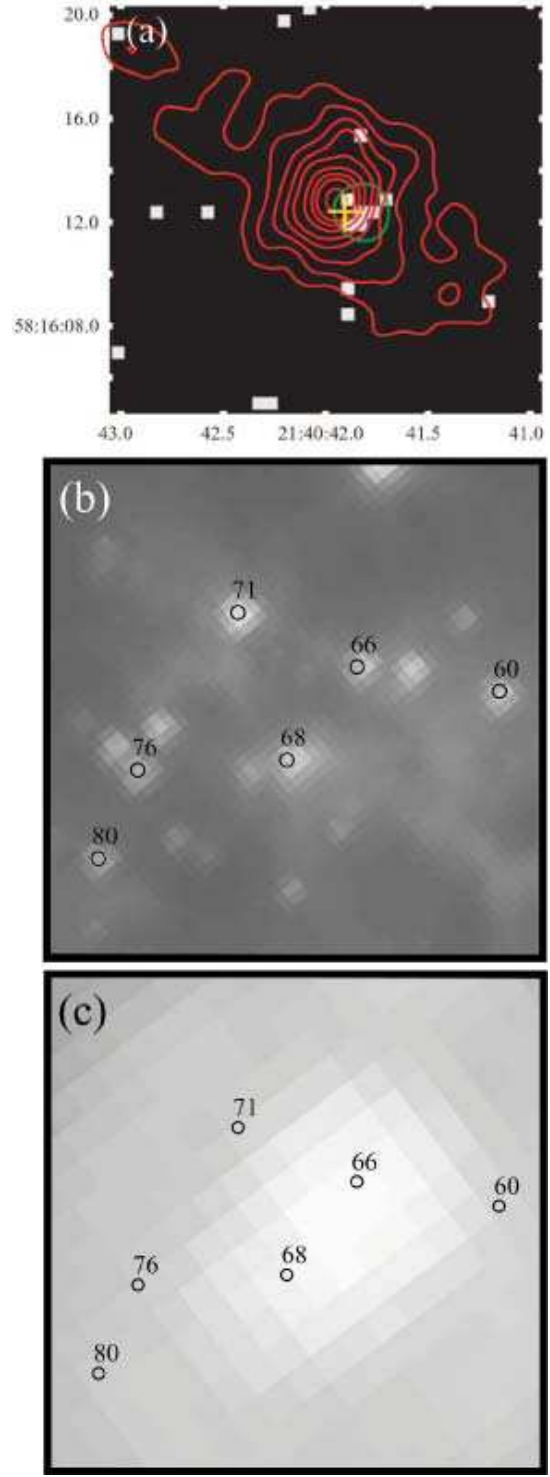}
\caption{Multiwavelength views of the region around protostar
IRAS~21391+5802. Panel (a) shows a $17\arcsec \times 16\arcsec$
closeup of the {\it Chandra} image with a pixel size of
0.5\arcsec.  The green circle contains the 8 extracted counts for
source \#66.  The X-ray position is consistent with the radio
continuum source VLA~2 (yellow $+$) and the 1.2 mm source BIMA~2
(red contours) obtained by \citet{Beltran02}. Panels (b) and (c)
show $1\arcmin \times 1\arcmin$ $Spitzer$ images in the 3.6 $\mu$m
IRAC and 70 $\mu$m MIPS bands, respectively. The six $Chandra$
protostars are labelled. \label{vla_bima_xray_fig}}
\end{figure}

Source \#66 in our $Chandra$ image is a faint, extremely hard
X-ray source within $0.5\arcsec$ of BIMA~2 = VLA~2 (Figure
\ref{vla_bima_xray_fig}$a$). Another X-ray source \#68 is within
$1\arcsec$ of BIMA~3 = VLA~3 \footnote{X-ray source \#60 lies
$3.2\arcsec$ north of VLA~1 and $5.0\arcsec$ northwest of BIMA~1.
This offset is too large to be considered a reliable coincidence.
Instead, we associate \#60 with NIR \#2 of \citet{Nisini01} with
bright MIR emission in the $Spitzer$ data lying within
0.8\arcsec\/ of the Chandra position.}. The X-ray spectrum of
\#66, though based on only 8 extracted photons, is very unusual:
it has the highest median energy of all sources in the field
($MedE=6.0$~keV; see Table \ref{tbl_xray}, Figures
\ref{phot_arriv_fig} and \ref{c_vs_mede_fig}).  The inferred
absorbing column density is $\log N_H \sim 24.0$ cm$^{-2}$,
corresponding to $A_V \sim 500$ mag.  This is one of the most
heavily absorbed sources ever seen with the $Chandra$ observatory;
for example, its $MedE$ is similar to the most embedded protostar
in OMC-1 behind the Orion Nebula, COUP source \# 632
\citep{Grosso05}. Adopting this $N_H$ value, the inferred
intrinsic hard band luminosity corrected for absorption is $\log
L_{h,c} \sim 31.3 \pm 0.4$ ergs~s$^{-1}$, by far the most luminous
X-ray source in the IC~1396N globule.

Examination of the photon arrival times shows that seven out of
eight X-ray photons detected from source \#66 arrived within the
first half of the $Chandra$ exposure (Figure
\ref{phot_arriv_fig}).  A Kolmogorov-Smirnov test gives only a
hint of possible variability. To evaluate this independently, we
simulated arrival times for 8 photons assuming constant emission
and tabulated the fraction of samples with splits \{7,1\} and
\{8,0\} across the mid-point of the observing interval. The
probability for a constant source to produce such an imbalanced
arrival pattern is 3\%.  It is thus likely that IRAS~21391+5802
was seen during the decay of a magnetic reconnection flare during
our $Chandra$ observation.

The $Spitzer$ 70 $\mu$m image (Figure \ref{vla_bima_xray_fig}$c$)
establishes the high far-infrared luminosities, and thus extreme
youth, of the protostars \#66 (BIMA~2) and \#68 (BIMA~3).  The
other four X-ray protostars show 3.6 $\mu$m emission (Figure
\ref{vla_bima_xray_fig}$b$) but are not as bright in 70 $\mu$m as
\# 66 and 68.  Other measures all indicate that \#66 is the
youngest object in the globule: extreme colors in the
$[3.6]-[4.5]-[5.8]$ color-color diagram (Figure
\ref{nir_mir_fig}$d$), the extraordinary X-ray absorption (above),
the high bolometric luminosity, strong submillimeter emission,
thermal centimeter emission, and a collimated molecular outflow
(\S \ref{IRAS_subsection}). In their detailed study of
IRAS~21391+5802, \citet{Beltran02} conclude that BIMA~2 is the
youngest object in the globule and is an intermediate-mass
protostar with circumstellar mass $\sim 5$~M$_\odot$.

\subsection{Comparison with other X-ray Emitting Protostars
\label{xrays_protostars_section}}

We seek here to understand how the IC~1396N X-ray protostars, and
particularly \#66 = IRAS~21391+5802, fit in the larger context of
X-rays from protostars.  The critical astrophysical question is
whether X-rays are present in Class 0 stars at the very onset of
star formation when collimated outflows begin.  This is required
if X-rays are the principal ionization source at the base of
protostellar molecular outflows
\citep{Shang02,FerroFontan03,Shang04}.

Tables \ref{tbl_xrayprotostars} and \ref{tbl_classi_irprop}
summarize a literature search of young protostars observed at high
sensitivity in X-rays.  Quantitative evaluation of X-ray
luminosities and upper limits is difficult to obtain, as the
values are strongly dependent on the intervening column density
which is often poorly determined. We thus give protostellar
distances, X-ray exposure times, and a binary indicator whether
the source was detected. Pre-$Chandra$/$XMM$ observations are much
less sensitive and are ignored.  Table \ref{tbl_xrayprotostars}a
gives results for 32 $confirmed$ very young (mostly Class~0)
protostars in a variety of star forming regions, including
IC~1396N BIMA~2 discussed here. Tables \ref{tbl_xrayprotostars}b
and \ref{tbl_classi_irprop} focus on well-characterized samples of
the nearby Ophiuchus and Serpens cloud older (Class~I) protostars.
We list $\alpha_{2-14}$, the spectral index between $6.7\mu$m and
$14.3\mu$m from $ISOCAM$ studies, and $K-$band magnitude from NIR
studies.

\begin{figure}
\centering
\includegraphics[angle=0.,width=7.0in]{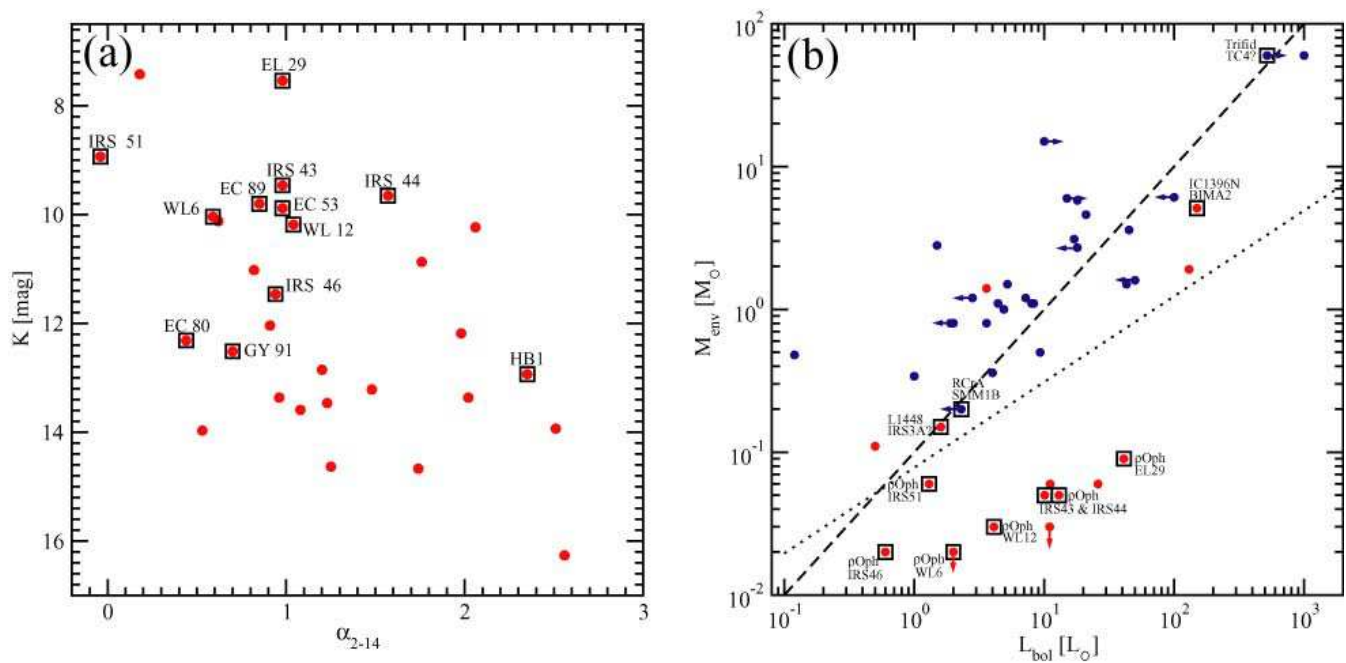}
\caption{X-ray detectability of protostars. X-ray detected objects
are outlined with $\Box$ and labeled.  (a) $K$-band magnitudes
$vs.$ IR spectral index for Class I protostars observed with
$Chandra$ and $XMM$ in the Ophiuchus and Serpens clouds (Tables
\ref{tbl_xrayprotostars}b and\ref{tbl_classi_irprop}). $K$
magnitudes for objects in Serpens are adjusted to match the
distance of Ophiuchus. (b) $M_{env}$ $vs.$ $L_{bol}$ Class~0 and I
protostars observed with $Chandra$ and $XMM$ (Table
\ref{tbl_xrayprotostars}a). The symbol colors, dashed and dotted
lines mark boundaries between Class~0 and Class~I stages; see text
for details. \label{protostars_menv_lbol_fig}}
\end{figure}

Figure \ref{protostars_menv_lbol_fig}$a$ shows the X-ray
detectability of Class I protostars in the Ophiuchus and Serpens
clouds as a function of $K$ band luminosity (a surrogate for
bolometric luminosity and mass) and infrared spectral index
\citep[a surrogate for disk mass and extinction and thus for
evolutionary stage,][]{Andre00}. Here, the $K-$band magnitudes for
Serpens stars are adjusted to match the distance of the Ophiuchus
cloud. The diagram shows that twelve out of 34 Class I protostars
have been detected in X-rays (symbols with boxes).  Detections are
preferentially protostars with higher stellar mass and later
evolutionary stage.

Figure \ref{protostars_menv_lbol_fig}$b$ displays X-ray detections
in a plot with three different discriminants between Class~0 and
Class~I systems.  The blue $vs.$ red symbols show the
classification in the published literature (Table
\ref{tbl_xrayprotostars}). The dashed line shows the ratio
$M_{env}/L_{bol} = 0.1$ in M$_\odot$/L$_\odot$ units, which is
roughly equivalent to the criterion $L_{smm}/L_{bol}=0.005$ of
\citet{Andre00}.  The dotted line marks the locus where half of
the mass of the initial pre-stellar condensations have been
accreted in the evolutionary models of \citet{Andre00}.

This diagram does not show the clear link between X-rays and
bolometric luminosity seen in Figure
\ref{protostars_menv_lbol_fig}$a$, but does show the association
with evolutionary stage.  Most of the X-ray detections have low
M$_{env}$ and lie clearly in the Class~I regime.  Four X-ray
detected stars lie along the poorly-defined Class 0/I boundary:
IRS~3A in Lynds 1448 \citep{Tsujimoto05}, SMM~1B (= IRS~7B) in R
Corona Australis \citep[R CrA,][]{Hamaguchi05}, TC~4 in the Trifid
Nebula \citep{Rho04}, and BIMA~2 (= IRAS~21391+5802) in the
IC~1396N globule discussed here\footnote{We omit here the early
report of Class~0 X-ray source slightly displaced from a binary
system of Class~I protostars in the OMC 2/3 region
\citep{Tsuboi01,Tsujimoto02b}.  NIR and MIR photometry indicate
these stars have Class~I SEDs with $A_V > 50$ mag
\citep{Nielbock03,Tsujimoto04}.  The X-ray source lies $\sim
1$\arcsec\/ from one component along an ionized jet, and thus may
arise from strong shocks in the outflow (Tsujimoto et al.\ 2004).
There is no evidence for a Class~0 source at the location of the
X-ray source, although this possibility is not firmly excluded.}.

Only one of the three previous reports appears reliable.  The
X-ray source attributed to TC~4 lies at the edge of the ACIS field
were point spread function is degraded, and its reported position
is $12\arcsec$ from the millimeter dust continuum source TC~4 and
could also be associated with more evolved 2MASS sources
\citep{Lefloch00, Rho04}. The X-ray source associated with the
very young low-mass Class~I protostar IRS~3A in L~1448 has only
three photons, although they arrived during first 5 hours of the
19~hour $Chandra$ exposure suggesting a flare
\citep{Ciardi03,Tsujimoto05}.  The detection of a Class 0/I
protostar in R CrA appears reliable. The X-ray source has dozens
of $Chandra$ photons and a thousand $XMM$ photons \citep[][also
see footnote to Table \ref{tbl_xrayprotostars}]{Hamaguchi05}.  The
X-ray position precisely coincides the deeply embedded radio
source VLA~10B \citep{Feigelson98} which \citet{Nutter05}
identifies with the submillimeter source SMM~1B. There are no
other nearby NIR or MIR sources, and the SED clearly denotes a
Class~0 protostar.

In this context, our report of 8 X-ray photons from BIMA~2 in the
IC~1396N globule represents the second excellent case for X-rays
from a Class 0/I star. Its borderline Class~0/I status is based on
its SED and outflow characteristics, as argued in detail by
\citet{Beltran02}. The ultra-hard X-ray spectrum guarantees that
the $Chandra$ events are not from instrumental background or
un-embedded star.  The positions at millimeter, radio, MIR, FIR
and X-ray wavebands agree to better than an arcsecond.  As
discussed in \S \ref{absorption_subsection}, the X-ray absorption
is attributed to the local infalling envelope of high mass
$M_{env}$. The high X-ray luminosity reasonably scales with its
high mass, given the correlation between stellar mass and X-ray
emission seen in Class I/II/III systems (Figures
\ref{lmag_vs_lhc_fig} and \ref{protostars_menv_lbol_fig}$a$). We
can reasonably guess that the underlying star has mass
$2-5$~M$_\odot$, which is consistent with $L_x \sim 10^{31}$
erg~s$^{-1}$.  The X-ray source would not have been detected in
our short exposure has the protostar had a more typical mass $M <
1$~M$_\odot$ and X-ray luminosity $L_x \sim 10^{28-30}$~erg
s$^{-1}$.

\section{Implications for Triggered Star Formation \label{trigger_section}}

The results of this study constitute strong endorsement for models
of triggered star formation in CGs embedded in giant HII regions.
Our basic findings are summarized in the maps of the IC~1396N
globule: Figure \ref{spat_distrib2_fig}$b$ shows the 25 X-ray
stars with respect to the infrared dust emission and Figure
\ref{nisini_fig}$a$ adds the red NIR stars of \citet{Nisini01}.
The stellar members are not uniformly distributed across the
globule, but rather are concentrated in an elongated pattern $\sim
0.8$~pc long and $\sim 0.15$~pc wide oriented towards the ionizing
star HD~206267.  This spatial distribution immediately suggests a
physical relationship between star formation and the ionization
front in the globule.

The stellar $Chandra$ grouping shows a distinct age gradient. Two of
the three Class~III stars (yellow triangles in the maps) lie nearly
0.5~pc outside the current globule boundary towards the ionizing
star. Most of the Class~II stars (red circles) are superposed on the
bright rim, again in front of the molecular globule.  The Class~0/I
protostars (blue squares) lie in a tight concentration of $\sim 0.2
\times 0.2$~pc$^2$ embedded in the molecular material.  The
\citet{Nisini01} stars lie furthest from the ionization front but, as
discussed in \S~\ref{nisini_cluster_subsection}, many of these stars
may not be members of the cluster.  This age gradient directly
supports a causal relationship between the progressive evaporation of
the globule by the ionization front and star formation.

These two phenomena -- an elongated spatial distribution and an age
gradient oriented towards the exciting star --  have been found in
other irradiated bright-rimmed CGs by Sugitani and colleagues. They
established catalogs of several dozen bright-rimmed globules with
embedded $IRAS$ sources or T~Tauri stars \citep{Sugitani91,
Sugitani94, Sugitani00}.  These include other globules in the giant
IC~1396 HII region \citep{Nakano89,Sugitani97}.  Age gradients, with
the youngest stars deeply embedded and older stars aligned towards
the ionizing star as we see in IC~1396N, were reported in a number of
cases including the well-known Eagle Nebula, M~16
\citep{Sugitani95,Sugitani02,Matsuyanagi06}. They frequently find "a
small cluster of red stars that is elongated along the axis of the
clouds, toward the tip of the BRC [bright rimmed cloud] and extending
farther out", arising from a process they call "small-scale
sequential star formation" \citep{Ogura02}.  This is supported by the
spatial distribution of young stars in the L~1616 cloud near the
Orion complex  \citep{Stanke02} and distribution of H$_2$O masers in
several irradiated clouds \citep{Healy04}. Our $Chandra$ findings
have an advantage in that older disk-free PMS stars can be sampled,
and many younger highly obscured protostars can be discovered.
Previous studies were largely restricted to very young often
spatially unresolved systems with millimeter and infrared excesses
from dusty disks and accreting T~Tauri stars with strong optical
emission lines.

Based on the triggered star formation concept, we can make a some
simple model-independent estimates of the triggering process. If we
assume that the T-Tauri population (Class~II and Class~III stars),
lying $0.3-0.5$~pc from the embedded population (Figure
\ref{spat_distrib2_fig}$b$), have ages $\sim 0.5-1$~Myr, and the
embedded Class 0/I protostars have typical ages $\sim 0.1$~Myr, we
find that the triggering shock propagated at a velocity $\sim 6 \pm
3$~pc/1~Myr or 0.6~km~s$^{-1}$.  At this rate, the globule should be
completely shocked in another $\sim 1$~Myr.  The star formation
efficiency for the T-Tauri population can be roughly estimated to be
$\sim 4$\% with $\sim 15$~M$_\odot$ in stars (\S
\ref{spectra_subsection}) and $\sim 400$~M$_\odot$ in molecular gas
(\S \ref{introduction_section}). Considering that the original
globule was substantially larger and more massive than it appears
today, the efficiency is probably below 4\%, perhaps $1-2$\%. This
suggests that the total population of second-generation triggered
stars is unlikely to exceed the population of the first-generation
cluster which produced the HII region.

A remarkable aspect of the $Chandra$ IC~1396N findings is the
discovery of 6 X-ray emitting protostars in an observation $\sim
50-70$ times less sensitive than the observations of Ophiuchus and
Serpens protostar clouds discussed in \S
\ref{xrays_protostars_section}. Number of X-ray detected
protostars in IC~1396N (6 protostars) is comparable with that of
Ophiuchus (8) and Serpens (4) observations, but their intrinsic
X-ray luminosities on average are above those in Ophiuchus and
Serpens protostar, implying higher mass objects, despite that the
estimated total population of previous generation of stars
(T-Tauri population) in IC~1396N is only $\sim 30$, several times
smaller than the clusters found in the Ophiuchus and Serpens
clouds. Two explanations suggest themselves. First, a population
of still-unidentified lower luminosity protostars may be present,
implying an extremely high ratio of Class 0/I $vs.$ Class II/III
systems compared to most embedded stellar clusters. Within the
assumption of a similar IMF shape for both protostellar and
T-Tauri populations, the former would actually outnumber the
latter resulting in an increase of the star formation efficiency
in the globule over time. Second, the process that triggered
IC~1396N globule may produce a non-standard IMF, preferentially
intermediate-mass stars rather than lower mass stars.
\citet{Sugitani91} suggested this would be a natural consequence
of radiation-driven implosion mechanism in CGs. In this case we
can estimate only a lower limit on the total mass of the detected
protostellar population of $\ga 9$~M$_{\odot}$, which would yield
a star formation efficiency of $\sim 2$\% or higher. Either of
these explanations indicates that IC~1396N, and possibly other
CGs, have unusual protostellar populations.

The relatively high number of X-ray protostars found in IC~1396N
indicates an early evolutionary stage of the globule.  Analytic
theory of the radiation driven implosion (RDI) model for triggered
star formation by HII regions shows that the process occurs in two
stages: the early collapse phase during which star formation is
induced, followed by a stationary regime when a globule has
developed the characteristic cometary structure
\citep{Bertoldi89,Bertoldi90}. Two-dimensional hydrodynamical
simulations show that the collapse stage occurs rapidly in  $\sim
0.1$~Myr ($\sim 10\%$ of the cloud's lifetime) with the first
compression triggering early star formation \citep{Lefloch94}.
Three-dimensional calculations indicate that the initial burst of
star formation may be delayed by $\sim 0.3$~Myr after the first
compression \citep{KesselDeynet03,Miao06}.

The calculated geometry of the RDI compressed cloud closely resembles
the spatial distribution of $Chandra$ stars in IC~1396N. The initial
passage of the ionization front is followed by several expansions and
re-compressions producing double-peaked molecular line profiles. The
second, cometary phase with the globule in a quasi-hydrostatic
equilibrium, has no conspicuous spectroscopic signatures, and lasts
longer (from a few $10^5$~yr to a few Myr, $\sim 90\%$ of the cloud's
lifetime). CO maps of the southern part of the IC~1396N globule show
blue- and red-shifted emission suggesting that IC~1396N is in the
transient early phase of expansions and re-compressions
\citep{Codella01}.

The collect-and-collapse model for triggered star formation has
been less well-studied.  In the hydrodynamical calculation of a
dense CG in the vicinity of an O star, the shock front generally
moves a bit faster than the ionization front.  The density spikes
between the fronts, and gravitational collapse can proceed in this
travelling thin shell \citep{Hosokawa06}.  This would produce a
spatial gradient in star ages similar to what we find in IC~1396N.
However, the calculated velocity of the front propagation into the
globule is typically several km~s$^{-1}$, which is an order of
magnitude faster than our estimate of $\sim 0.6 \pm
0.3$~km~s$^{-1}$ above.

\section{Conclusions \label{conclusion_section}}

We report results from a 30~ks $Chandra$ study of the stellar
population in the cometary globule IC~1396N, previously recognized
as a site of star formation triggered by an HII region expanding
into an inhomogeneous molecular cloud.  Of the 117 X-ray sources
in the $17\arcmin \times 17\arcmin$\/ field, 25 appear associated
with young stars formed in the globule.  Although relatively
bright ($10<K<13.5$ mag), most of these were previously
unidentified due to confusion with unrelated Galactic stars. X-ray
selection is very effective in isolating PMS members from old
field stars, and the extragalactic contaminants can be removed
with some reliability. About $50-60$ of the other sources in the
X-ray field are likely members of the 3~Myr old open cluster
Trumpler~37 that is dispersed throughout the large IC~1396 HII
region.

2MASS NIR and $Spitzer$ MIR photometry permit evaluation of the
disk properties of the 25 X-ray stars.  Using the standard PMS
evolutionary classification, we find: 1 transitional Class~0/I
protostar, 5 Class~I protostars, 1 transitional Class~I/II star,
15 Class~II classical T Tauri stars, and 3 Class~III weak-lined T
Tauri stars.  This is clearly a younger population than many
X-ray-studied stellar clusters in nearby molecular clouds (e.g.,
Orion Nebula, Ophiuchus, Chamaeleon) where the Class~III
population outnumbers the younger systems.

From NIR photometry, the T-Tauri (lightly obscured) X-ray
population in the globule have estimated masses between 0.2 and
2~M$_\odot$ and typical absorptions of $3-4$ visual magnitudes.
Intrinsic X-ray luminosities range from $10^{28.5}$ to $10^{30.0}$
erg~s$^{-1}$ in the hard $2-8$ keV band with spectral and
variability characteristics typical of low-mass PMS stars. Using
the established relationship between X-ray luminosity and mass to
evaluate the population falling below our X-ray detection limit,
we estimate that the total T-Tauri population of the globule is
$\sim 30$ stars with an integrated mass $\sim 15$~M$_\odot$. This
yields a star formation efficiency of $1 - 4\%$ depending on the
original mass of the globule.

A spatial gradient in stellar ages is discovered with the protostars
deeply embedded in the IC~1396N globule, the Class~II systems
congregating in the bright ionized rim, and the older Class~III in
the outer rim. The stars lie in a line pointing towards the ionizing
O star HD~206267. This independently supports linear gradients of PMS
stars found by Sugitani and colleagues in several other cometary
globules, and confirms the basic hypothesis of triggering by passage
of HII region shocks into the molecular globule. Our estimated
velocity of shock front proceeding into the globule of $\sim
0.6$~km~s$^{-1}$ is an order of magnitude smaller than a typical
value of several km~s$^{-1}$ used in ``collect-and-collapse'' models
for triggered star formation.

The discovery of relatively high numbers of quite massive, X-ray
luminous protostars in IC~1396N globule from a short $Chandra$
exposure suggests that either a population of still-unidentified
lower mass protostars is present, implying an unusually high ratio
of Class I/0 $vs.$ Class II/III systems, or the triggering process
produces a non-standard IMF biased towards higher mass stars. The
X-ray protostellar population reveals the very young age ($\sim
0.1-0.3$ Myr) of the globule, which agrees with the gas dynamics
inferred from CO line shapes. This all supports the RDI model of
triggered star formation in IC~1396N.

We establish $Chandra$ source \#66 (CXOU J214041.81$+$581612.3) as
one of the youngest objects ever detected in the X-ray band.  It
lies within 0.5\arcsec\/ of the luminous far-infrared/millimeter
source IRAS~21391+5802 = BIMA~2, classified as a Class 0-I
borderline protostar.  In a novel comparison of X-ray absorption
with MIR colors, we demonstrate that the X-ray absorption of
protostellar objects should be attributed to the local molecular
gas in their infalling envelopes, rather than from ambient
molecular cloud material.

\acknowledgements We thank Philippe Andr\'e for correspondence on
the BIMA~2 protostar, Leisa Townsley for many discussions on
X-rays from star forming regions, Kevin Luhman for his useful
suggestions on infrared data analysis, and the anonymous referee
for his time and many useful comments that improved this work.
This work was supported by the $Chandra$ ACIS Team (G. Garmire,
PI) through NASA contract NAS8-38252. This publication makes use
of data products from the Two Micron All Sky Survey (a joint
project of the University of Massachusetts and the Infrared
Processing and Analysis Center/California Institute of Technology,
funded by NASA and NSF), archival data obtained with the Spitzer
Space Telescope (operated by the Jet Propulsion Laboratory,
California Institute of Technology under a contract with NASA),
and the SIMBAD database (operated at CDS, Strasbourg, France).

\clearpage

\begin{deluxetable}{rcrrrrrrrrrrrrcccccc}
\centering \rotate \tabletypesize{\tiny} \tablewidth{0pt}
\tablecolumns{20}

\tablecaption{X-ray Sources in the IC~1396N Field \label{tbl_bsp}}

\tablehead{ \multicolumn{2}{c}{Source} &
  &
\multicolumn{3}{c}{Position} &
  &
\multicolumn{5}{c}{Extracted Counts} &
  &
\multicolumn{5}{c}{Characteristics}  &
  & \\
\cline{1-2} \cline{4-6} \cline{8-12} \cline{14-18}

\colhead{Seq} & \colhead{CXOU~J} &
  &
\colhead{$\alpha_{\rm J2000}$} & \colhead{$\delta_{\rm J2000}$} &
\colhead{$\theta$} &
  &
\colhead{Net} & \colhead{$\Delta$Net} & \colhead{Bkgd} &
\colhead{Net} & \colhead{PSF} &
  &
\colhead{Signif} & \colhead{Anom\tablenotemark{a}} & \colhead{Var\tablenotemark{b}} & \colhead{EffExp} & \colhead{Med E} & & \colhead{2MASS} \\

\colhead{\#} & \colhead{} &
  &
\colhead{(deg)} & \colhead{(deg)} & \colhead{(\arcmin)} &
  &
\colhead{Full} & \colhead{Full} & \colhead{Full} & \colhead{Hard}
& \colhead{Frac} &
  &
\colhead{} & \colhead{} & \colhead{} & \colhead{(ks)} &
\colhead{(keV)} & & \colhead{}\\

\colhead{(1)} & \colhead{(2)} &
  &
\colhead{(3)} & \colhead{(4)} & \colhead{(5)} &
  &
\colhead{(6)} & \colhead{(7)} & \colhead{(8)} & \colhead{(9)} &
\colhead{(10)} &
  &
\colhead{(11)} & \colhead{(12)} & \colhead{(13)} & \colhead{(14)}
& \colhead{(15)} & & \colhead{(16)} }

\startdata
1 & 213926.70$+$581240.2 & & 324.861261 & 58.211189 & 10.30 & & 93.6 & 10.6 & 8.4 & 18.1 & 0.91 & & 8.4 & g... & \nodata & 19.7 & 1.34 & & 21392667$+$5812405 \\
2 & 213936.78$+$581827.1 & & 324.903254 & 58.307541 & 8.78 & & 21.4 & 5.9 & 7.6 & 6.6 & 0.90 & & 3.3 & .... & a & 22.5 & 1.47 & & 21393680$+$5818272 \\
3 & 213940.32$+$581319.2 & & 324.918020 & 58.222026 & 8.39 & & 23.4 & 5.9 & 5.6 & 14.1 & 0.90 & & 3.6 & .... & a & 23.1 & 2.28 & & \nodata \\
4 & 213940.82$+$581816.8 & & 324.920110 & 58.304670 & 8.22 & & 61.1 & 8.7 & 4.9 & 28.1 & 0.90 & & 6.7 & g... & \nodata & 22.0 & 1.92 & & \nodata \\
5 & 213943.92$+$581326.0 & & 324.933015 & 58.223893 & 7.91 & & 8.4 & 4.2 & 4.6 & 10.0 & 0.90 & & 1.8 & .... & a & 23.5 & 4.52 & & \nodata \\
6 & 213944.54$+$581443.9 & & 324.935613 & 58.245555 & 7.51 & & 55.3 & 8.1 & 1.7 & 11.9 & 0.78 & & 6.4 & .... & a & 24.2 & 1.28 & & 21394443$+$5814438 \\
7 & 213945.07$+$581450.5 & & 324.937818 & 58.247380 & 7.42 & & 501.1 & 23.0 & 2.9 & 138.1 & 0.89 & & 21.3 & .... & a & 24.3 & 1.53 & & 21394504$+$5814504 \\
8 & 213945.63$+$581616.8 & & 324.940141 & 58.271361 & 7.27 & & 9.1 & 4.0 & 2.9 & 2.0 & 0.89 & & 2.0 & .... & b & 24.2 & 1.74 & & 21394555$+$5816172 \\
9 & 213948.64$+$582030.3 & & 324.952706 & 58.341754 & 8.22 & & 7.7 & 4.0 & 4.3 & 4.6 & 0.90 & & 1.7 & .... & a & 23.8 & 2.26 & & \nodata \\
10 & 213949.34$+$581616.5 & & 324.955584 & 58.271271 & 6.78 & & 17.6 & 5.0 & 2.4 & 11.2 & 0.89 & & 3.2 & .... & a & 25.1 & 2.34 & & \nodata \\
11 & 213950.12$+$581149.9 & & 324.958850 & 58.197203 & 7.86 & & 30.2 & 6.4 & 3.8 & 9.7 & 0.90 & & 4.4 & .... & c & 22.1 & 1.48 & & 21395001$+$5811503 \\
12 & 213951.69$+$581112.5 & & 324.965407 & 58.186833 & 8.04 & & 17.4 & 5.1 & 3.6 & 0.0 & 0.89 & & 3.1 & .... & a & 23.1 & 1.38 & & \nodata \\
13 & 213952.85$+$582003.6 & & 324.970226 & 58.334339 & 7.51 & & 12.1 & 4.5 & 3.9 & 7.5 & 0.90 & & 2.4 & .... & a & 24.7 & 2.46 & & 21395268$+$5820047 \\
14 & 213953.43$+$582031.6 & & 324.972640 & 58.342138 & 7.72 & & 10.5 & 4.3 & 3.5 & 7.6 & 0.90 & & 2.2 & .... & a & 24.4 & 2.83 & & 21395324$+$5820323 \\
15 & 213953.81$+$581240.6 & & 324.974214 & 58.211304 & 7.01 & & 15.8 & 4.8 & 2.2 & 7.7 & 0.89 & & 3.0 & .... & a & 22.9 & 2.50 & & \nodata \\
16 & 213954.64$+$581618.5 & & 324.977675 & 58.271818 & 6.09 & & 171.3 & 13.7 & 1.7 & 27.9 & 0.90 & & 12.1 & .... & a & 25.8 & 1.32 & & 21395459$+$5816185 \\
17 & 213954.84$+$581200.6 & & 324.978520 & 58.200188 & 7.24 & & 13.1 & 4.5 & 2.9 & 2.2 & 0.90 & & 2.6 & .... & a & 24.2 & 1.28 & & 21395480$+$5812009 \\
18 & 213956.40$+$581347.4 & & 324.985029 & 58.229838 & 6.24 & & 207.4 & 15.0 & 1.6 & 61.9 & 0.90 & & 13.4 & .... & a & 24.0 & 1.61 & & 21395635$+$5813475 \\
19 & 213956.55$+$581113.3 & & 324.985637 & 58.187052 & 7.52 & & 105.2 & 10.9 & 2.8 & 27.0 & 0.90 & & 9.2 & .... & a & 23.7 & 1.50 & & 21395651$+$5811136 \\
20 & 213958.46$+$581214.8 & & 324.993605 & 58.204115 & 6.71 & & 76.9 & 9.4 & 2.1 & 21.5 & 0.89 & & 7.7 & .... & a & 24.8 & 1.53 & & 21395844$+$5812147 \\
21 & 214002.10$+$581116.1 & & 325.008751 & 58.187825 & 6.94 & & 19.0 & 5.1 & 2.0 & 3.4 & 0.90 & & 3.3 & .... & a & 23.4 & 1.29 & & 21400217$+$5811164 \\
22 & 214002.44$+$581224.3 & & 325.010206 & 58.206756 & 6.19 & & 10.3 & 4.0 & 1.7 & 0.7 & 0.90 & & 2.2 & .... & a & 25.2 & 0.96 & & 21400244$+$5812237 \\
23 & 214002.62$+$581035.7 & & 325.010947 & 58.176585 & 7.37 & & 6.6 & 3.5 & 2.4 & 4.4 & 0.89 & & 1.6 & .... & b & 23.7 & 2.18 & & \nodata \\
24 & 214002.93$+$581509.7 & & 325.012247 & 58.252722 & 5.05 & & 5.5 & 3.0 & 0.5 & 0.6 & 0.89 & & 1.5 & .... & a & 24.6 & 1.06 & & 21400298$+$5815097 \\
25 & 214003.75$+$581149.0 & & 325.015636 & 58.196964 & 6.41 & & 5.6 & 3.2 & 1.4 & 2.0 & 0.90 & & 1.5 & .... & a & 24.0 & 1.30 & & 21400356$+$5811478 \\
26 & 214004.12$+$581456.6 & & 325.017202 & 58.249077 & 4.94 & & 5.4 & 3.0 & 0.6 & 4.6 & 0.90 & & 1.5 & .... & a & 26.9 & 5.82 & & \nodata \\
27 & 214006.06$+$581757.1 & & 325.025262 & 58.299214 & 4.98 & & 5.5 & 3.0 & 0.5 & 2.7 & 0.90 & & 1.5 & g... & \nodata & 18.8 & 3.07 & & \nodata \\
28 & 214006.31$+$581823.4 & & 325.026314 & 58.306511 & 5.14 & & 9.2 & 3.7 & 0.8 & 2.5 & 0.89 & & 2.2 & .... & a & 24.6 & 1.55 & & 21400633$+$5818233 \\
29 & 214009.00$+$581409.6 & & 325.037513 & 58.236007 & 4.57 & & 5.4 & 3.0 & 0.6 & 0.0 & 0.89 & & 1.5 & .... & a & 27.1 & 1.06 & & 21400909$+$5814094 \\
30 & 214010.19$+$581640.7 & & 325.042494 & 58.277992 & 4.09 & & 25.5 & 5.6 & 0.5 & 5.7 & 0.90 & & 4.1 & .... & a & 27.8 & 1.31 & & 21401012$+$5816406 \\
31 & 214010.32$+$580900.1 & & 325.043019 & 58.150038 & 8.05 & & 37.5 & 6.8 & 2.5 & 10.3 & 0.89 & & 5.1 & g... & \nodata & 17.3 & 1.51 & & 21401031$+$5809002 \\
32 & 214010.46$+$581701.9 & & 325.043596 & 58.283865 & 4.13 & & 31.6 & 6.2 & 0.4 & 23.7 & 0.90 & & 4.7 & .... & a & 27.8 & 3.69 & & \nodata \\
33 & 214010.52$+$580926.9 & & 325.043838 & 58.157480 & 7.65 & & 48.1 & 7.7 & 2.9 & 7.0 & 0.90 & & 5.9 & .... & a & 22.6 & 1.35 & & 21401056$+$5809270 \\
34 & 214012.27$+$581349.2 & & 325.051148 & 58.230345 & 4.33 & & 3.5 & 2.5 & 0.5 & 0.7 & 0.89 & & 1.1 & .... & a & 27.2 & 1.70 & & 21401225$+$5813495 \\
35 & 214015.29$+$582156.5 & & 325.063726 & 58.365707 & 6.85 & & 10.1 & 4.0 & 1.9 & 9.7 & 0.89 & & 2.2 & .... & a & 23.3 & 3.45 & & \nodata \\
36 & 214016.08$+$581522.2 & & 325.067002 & 58.256174 & 3.31 & & 11.7 & 4.0 & 0.3 & 8.8 & 0.90 & & 2.6 & .... & a & 26.9 & 3.28 & & \nodata \\
37 & 214016.24$+$581208.0 & & 325.067700 & 58.202240 & 5.02 & & 5.2 & 3.0 & 0.8 & 0.4 & 0.89 & & 1.4 & .... & a & 26.1 & 1.47 & & \nodata \\
38 & 214017.35$+$581321.7 & & 325.072300 & 58.222706 & 4.05 & & 12.6 & 4.1 & 0.4 & 0.8 & 0.89 & & 2.7 & .... & a & 27.2 & 1.12 & & 21401728$+$5813217 \\
39 & 214020.37$+$581323.5 & & 325.084890 & 58.223211 & 3.73 & & 5.7 & 3.0 & 0.3 & 1.8 & 0.89 & & 1.6 & .... & c & 27.5 & 1.25 & & 21402034$+$5813235 \\
40 & 214026.88$+$582230.1 & & 325.112009 & 58.375036 & 6.78 & & 8.2 & 3.7 & 1.8 & 2.6 & 0.89 & & 1.9 & .... & a & 25.1 & 1.32 & & \nodata \\
41 & 214027.31$+$581421.1 & & 325.113817 & 58.239215 & 2.41 & & 10.8 & 3.8 & 0.2 & 3.9 & 0.90 & & 2.4 & .... & a & 28.4 & 1.47 & & 21402732$+$5814212 \\
42 & 214028.19$+$581905.7 & & 325.117497 & 58.318272 & 3.54 & & 4.7 & 2.8 & 0.3 & 3.8 & 0.89 & & 1.4 & .... & a & 27.1 & 3.35 & & \nodata \\
43 & 214028.65$+$582255.6 & & 325.119388 & 58.382120 & 7.14 & & 5.8 & 3.4 & 2.2 & 0.0 & 0.90 & & 1.5 & .... & a & 24.6 & 1.07 & & 21402856$+$5822556 \\
44 & 214028.66$+$582157.2 & & 325.119440 & 58.365905 & 6.19 & & 26.8 & 5.8 & 1.2 & 7.1 & 0.90 & & 4.2 & g... & \nodata & 23.9 & 1.09 & & 21402865$+$5821574 \\
45 & 214028.71$+$582011.6 & & 325.119664 & 58.336568 & 4.51 & & 45.6 & 7.3 & 0.4 & 10.7 & 0.89 & & 5.8 & .... & a & 25.8 & 1.44 & & 21402871$+$5820115 \\
46 & 214028.80$+$581754.6 & & 325.120028 & 58.298506 & 2.50 & & 5.8 & 3.0 & 0.2 & 5.8 & 0.89 & & 1.6 & .... & a & 29.2 & 4.84 & & \nodata \\
47 & 214029.69$+$581645.9 & & 325.123746 & 58.279437 & 1.67 & & 5.9 & 3.0 & 0.1 & 4.9 & 0.89 & & 1.6 & .... & a & 29.3 & 3.10 & & \nodata \\
48 & 214030.21$+$580935.5 & & 325.125915 & 58.159880 & 6.53 & & 292.4 & 17.7 & 1.6 & 73.0 & 0.89 & & 16.1 & .... & a & 24.8 & 1.45 & & 21403017$+$5809353 \\
49 & 214031.58$+$581755.2 & & 325.131589 & 58.298668 & 2.30 & & 26.8 & 5.7 & 0.2 & 6.8 & 0.89 & & 4.3 & .... & a & 29.3 & 1.61 & & 21403159$+$5817551 \\
50 & 214032.87$+$581744.2 & & 325.136960 & 58.295635 & 2.05 & & 21.8 & 5.2 & 0.2 & 16.9 & 0.89 & & 3.8 & .... & a & 29.4 & 3.34 & & \nodata \\
51 & 214035.42$+$581230.9 & & 325.147607 & 58.208591 & 3.53 & & 75.8 & 9.2 & 0.2 & 17.8 & 0.89 & & 7.8 & .... & a & 25.1 & 1.50 & & 21403538$+$5812308 \\
52 & 214035.64$+$582435.4 & & 325.148519 & 58.409857 & 8.65 & & 17.3 & 5.1 & 3.7 & 8.4 & 0.90 & & 3.1 & g... & \nodata & 21.7 & 1.89 & & \nodata \\
53 & 214036.57$+$581345.8 & & 325.152388 & 58.229404 & 2.28 & & 28.8 & 5.9 & 0.2 & 8.9 & 0.90 & & 4.5 & .... & a & 28.7 & 1.45 & & 21403655$+$5813458 \\
54 & 214036.59$+$581133.0 & & 325.152461 & 58.192524 & 4.46 & & 36.4 & 6.6 & 0.6 & 11.6 & 0.89 & & 5.1 & .... & a & 27.0 & 1.45 & & 21403659$+$5811330 \\
55 & 214036.90$+$581437.9 & & 325.153771 & 58.243884 & 1.44 & & 10.8 & 3.8 & 0.2 & 1.9 & 0.90 & & 2.4 & .... & a & 29.4 & 1.41 & & 21403691$+$5814378 \\
56 & 214038.36$+$580726.9 & & 325.159836 & 58.124158 & 8.53 & & 25.9 & 5.9 & 3.1 & 3.9 & 0.90 & & 4.0 & .... & a & 20.7 & 1.42 & & 21403827$+$5807266 \\
57 & 214038.86$+$581922.4 & & 325.161934 & 58.322913 & 3.41 & & 4.7 & 2.8 & 0.3 & 4.8 & 0.90 & & 1.4 & .... & a & 28.0 & 3.04 & & \nodata \\
58 & 214038.98$+$581413.9 & & 325.162446 & 58.237204 & 1.76 & & 19.9 & 5.0 & 0.1 & 17.9 & 0.90 & & 3.6 & g... & \nodata & 23.4 & 3.48 & & \nodata \\
59 & 214039.17$+$580815.6 & & 325.163241 & 58.137689 & 7.72 & & 15.3 & 4.8 & 2.7 & 10.0 & 0.90 & & 2.9 & g... & \nodata & 21.9 & 3.47 & & \nodata \\
60 & 214039.62$+$581609.3 & & 325.165107 & 58.269272 & 0.24 & & 26.9 & 5.7 & 0.1 & 26.9 & 0.90 & & 4.3 & .... & a & 29.5 & 4.62 & & \nodata \\
61 & 214039.87$+$581834.8 & & 325.166135 & 58.309685 & 2.61 & & 113.9 & 11.2 & 0.1 & 94.9 & 0.89 & & 9.7 & .... & c & 26.6 & 3.54 & & 21403988$+$5818349 \\
62 & 214041.12$+$581359.0 & & 325.171337 & 58.233060 & 1.99 & & 71.8 & 9.0 & 0.2 & 14.9 & 0.90 & & 7.5 & .... & a & 27.4 & 1.45 & & 21404110$+$5813588 \\
63 & 214041.16$+$581511.2 & & 325.171534 & 58.253134 & 0.79 & & 46.9 & 7.4 & 0.1 & 29.9 & 0.90 & & 5.9 & .... & a & 29.4 & 2.42 & & 21404116$+$5815112 \\
64 & 214041.18$+$581201.4 & & 325.171594 & 58.200406 & 3.95 & & 6.6 & 3.2 & 0.4 & 0.0 & 0.89 & & 1.7 & .... & a & 26.3 & 1.03 & & 21404115$+$5812013 \\
65 & 214041.56$+$581425.5 & & 325.173203 & 58.240442 & 1.55 & & 8.9 & 3.5 & 0.1 & 2.9 & 0.90 & & 2.2 & g... & \nodata & 21.6 & 1.76 & & 21404156$+$5814255 \\
66 & 214041.81$+$581612.3 & & 325.174232 & 58.270097 & 0.26 & & 7.9 & 3.4 & 0.1 & 7.9 & 0.90 & & 2.0 & .... & a & 29.4 & 5.99 & & \nodata \\
67 & 214041.91$+$581523.1 & & 325.174664 & 58.256436 & 0.60 & & 14.9 & 4.4 & 0.1 & 7.9 & 0.90 & & 3.0 & .... & a & 26.7 & 2.28 & & 21404191$+$5815230 \\
68 & 214042.89$+$581601.0 & & 325.178727 & 58.266953 & 0.27 & & 4.9 & 2.8 & 0.1 & 3.9 & 0.90 & & 1.4 & .... & a & 29.3 & 5.61 & & \nodata \\
69 & 214043.47$+$581227.2 & & 325.181136 & 58.207582 & 3.54 & & 3.7 & 2.5 & 0.3 & 3.8 & 0.90 & & 1.2 & .... & a & 26.7 & 5.10 & & \nodata \\
70 & 214043.47$+$581559.7 & & 325.181139 & 58.266605 & 0.34 & & 25.9 & 5.6 & 0.1 & 22.9 & 0.90 & & 4.2 & .... & c & 29.3 & 3.28 & & \nodata \\
71 & 214043.64$+$581618.9 & & 325.181852 & 58.271932 & 0.50 & & 8.9 & 3.5 & 0.1 & 6.9 & 0.90 & & 2.1 & .... & a & 29.3 & 3.85 & & 21404365$+$5816190 \\
72 & 214044.34$+$581513.3 & & 325.184770 & 58.253699 & 0.88 & & 3.9 & 2.5 & 0.1 & 1.9 & 0.90 & & 1.2 & g... & \nodata & 21.8 & 1.39 & & 21404433$+$5815133 \\
73 & 214044.84$+$581605.1 & & 325.186858 & 58.268109 & 0.54 & & 15.9 & 4.5 & 0.1 & 9.9 & 0.90 & & 3.1 & .... & a & 29.3 & 2.20 & & 21404485$+$5816050 \\
74 & 214044.85$+$581503.4 & & 325.186883 & 58.250963 & 1.06 & & 39.9 & 6.8 & 0.1 & 11.9 & 0.90 & & 5.4 & .... & a & 27.5 & 1.61 & & 21404485$+$5815033 \\
75 & 214045.15$+$581950.5 & & 325.188148 & 58.330706 & 3.91 & & 32.7 & 6.3 & 0.3 & 21.8 & 0.90 & & 4.8 & .... & a & 25.9 & 2.45 & & 21404517$+$5819506 \\
76 & 214045.18$+$581559.8 & & 325.188264 & 58.266616 & 0.57 & & 52.9 & 7.8 & 0.1 & 52.9 & 0.90 & & 6.3 & .... & c & 29.2 & 4.23 & & \nodata \\
77 & 214045.51$+$581511.4 & & 325.189651 & 58.253167 & 0.99 & & 6.9 & 3.2 & 0.1 & 1.9 & 0.90 & & 1.8 & g... & \nodata & 22.9 & 1.74 & & 21404551$+$5815115 \\
78 & 214045.53$+$581602.9 & & 325.189715 & 58.267478 & 0.62 & & 9.9 & 3.7 & 0.1 & 4.9 & 0.90 & & 2.3 & .... & a & 29.2 & 1.91 & & 21404553$+$5816027 \\
79 & 214045.55$+$580846.9 & & 325.189810 & 58.146366 & 7.22 & & 3.5 & 3.0 & 2.5 & 2.2 & 0.90 & & 1.0 & .... & a & 23.5 & 5.13 & & \nodata \\
80 & 214045.79$+$581549.0 & & 325.190799 & 58.263625 & 0.67 & & 10.9 & 3.8 & 0.1 & 9.9 & 0.90 & & 2.5 & .... & a & 29.2 & 4.33 & & \nodata \\
81 & 214046.49$+$581523.2 & & 325.193746 & 58.256468 & 0.94 & & 11.9 & 4.0 & 0.1 & 4.0 & 0.90 & & 2.6 & g... & \nodata & 11.9 & 1.57 & & 21404649$+$5815230 \\
82 & 214046.89$+$581533.3 & & 325.195375 & 58.259270 & 0.90 & & 3.9 & 2.5 & 0.1 & 0.0 & 0.90 & & 1.2 & g... & \nodata & 22.0 & 1.42 & & 21404688$+$5815334 \\
83 & 214047.34$+$582025.5 & & 325.197279 & 58.340424 & 4.53 & & 5.4 & 3.0 & 0.6 & 4.6 & 0.90 & & 1.5 & .... & a & 25.2 & 5.18 & & \nodata \\
84 & 214047.63$+$581132.7 & & 325.198482 & 58.192424 & 4.52 & & 209.5 & 15.0 & 0.5 & 158.7 & 0.89 & & 13.5 & .... & b & 26.8 & 2.85 & & 21404762$+$5811328 \\
85 & 214048.03$+$581537.9 & & 325.200164 & 58.260540 & 1.00 & & 25.9 & 5.6 & 0.1 & 10.9 & 0.90 & & 4.2 & .... & a & 29.0 & 1.34 & & 21404803$+$5815376 \\
86 & 214048.20$+$581746.0 & & 325.200840 & 58.296123 & 2.04 & & 29.9 & 6.0 & 0.1 & 28.9 & 0.89 & & 4.6 & .... & a & 28.9 & 4.10 & & \nodata \\
87 & 214049.09$+$581709.3 & & 325.204549 & 58.285934 & 1.60 & & 17.9 & 4.8 & 0.1 & 8.9 & 0.89 & & 3.4 & .... & a & 28.9 & 1.93 & & 21404908$+$5817093 \\
88 & 214049.28$+$582055.3 & & 325.205362 & 58.348701 & 5.07 & & 4.3 & 2.8 & 0.7 & 1.5 & 0.89 & & 1.3 & .... & a & 24.2 & 1.89 & & \nodata \\
89 & 214050.33$+$580809.6 & & 325.209729 & 58.136021 & 7.91 & & 10.1 & 4.1 & 2.9 & 4.8 & 0.90 & & 2.1 & .... & a & 21.1 & 2.29 & & \nodata \\
90 & 214054.64$+$581204.7 & & 325.227680 & 58.201322 & 4.30 & & 12.5 & 4.1 & 0.5 & 8.7 & 0.89 & & 2.7 & .... & b & 26.7 & 2.31 & & \nodata \\
91 & 214055.43$+$582051.7 & & 325.230998 & 58.347704 & 5.25 & & 62.1 & 8.4 & 0.9 & 45.4 & 0.90 & & 6.9 & .... & b & 26.3 & 3.01 & & \nodata \\
92 & 214056.46$+$580604.2 & & 325.235256 & 58.101172 & 10.11 & & 7.5 & 4.6 & 8.5 & 0.2 & 0.90 & & 1.4 & g... & \nodata & 19.1 & 1.55 & & 21405643$+$5806037 \\
93 & 214057.24$+$582149.6 & & 325.238504 & 58.363801 & 6.24 & & 9.1 & 3.7 & 0.9 & 4.4 & 0.90 & & 2.1 & g... & \nodata & 15.4 & 1.98 & & \nodata \\
94 & 214059.72$+$581954.7 & & 325.248856 & 58.331876 & 4.65 & & 7.4 & 3.4 & 0.6 & 4.6 & 0.91 & & 1.9 & .... & a & 26.7 & 3.53 & & \nodata \\
95 & 214104.27$+$581130.7 & & 325.267823 & 58.191885 & 5.42 & & 6.1 & 3.2 & 0.9 & 4.4 & 0.91 & & 1.6 & .... & a & 25.5 & 3.22 & & \nodata \\
96 & 214104.85$+$582025.2 & & 325.270247 & 58.340351 & 5.45 & & 43.0 & 7.1 & 1.0 & 27.3 & 0.90 & & 5.6 & .... & a & 25.9 & 2.33 & & \nodata \\
97 & 214105.61$+$581049.2 & & 325.273405 & 58.180335 & 6.10 & & 60.9 & 8.4 & 1.1 & 47.2 & 0.90 & & 6.8 & g... & \nodata & 20.7 & 3.51 & & \nodata \\
98 & 214106.39$+$581630.3 & & 325.276634 & 58.275103 & 3.40 & & 8.7 & 3.5 & 0.3 & 1.8 & 0.90 & & 2.1 & .... & a & 27.3 & 1.82 & & 21410636$+$5816305 \\
99 & 214106.76$+$580646.9 & & 325.278188 & 58.113054 & 9.80 & & 50.1 & 8.0 & 5.9 & 8.9 & 0.90 & & 5.9 & .... & a & 20.0 & 1.44 & & 21410669$+$5806459 \\
100 & 214108.44$+$581635.6 & & 325.285194 & 58.276570 & 3.68 & & 5.7 & 3.0 & 0.3 & 0.0 & 0.90 & & 1.6 & .... & a & 27.1 & 1.45 & & 21410840$+$5816356 \\
101 & 214109.10$+$581049.9 & & 325.287955 & 58.180543 & 6.35 & & 136.5 & 12.3 & 1.5 & 35.9 & 0.90 & & 10.7 & .... & c & 24.4 & 1.63 & & 21410909$+$5810495 \\
102 & 214110.96$+$581622.9 & & 325.295679 & 58.273051 & 3.98 & & 9.6 & 3.7 & 0.4 & 8.7 & 0.90 & & 2.2 & .... & a & 26.8 & 3.00 & & 21411114$+$5816240 \\
103 & 214113.93$+$582002.0 & & 325.308067 & 58.333905 & 5.95 & & 10.9 & 4.0 & 1.1 & 10.1 & 0.90 & & 2.4 & .... & a & 25.2 & 3.72 & & \nodata \\
104 & 214114.63$+$581503.0 & & 325.310980 & 58.250850 & 4.54 & & 30.5 & 6.1 & 0.5 & 13.6 & 0.89 & & 4.6 & .... & a & 26.1 & 1.69 & & \nodata \\
105 & 214115.06$+$580652.0 & & 325.312787 & 58.114467 & 10.16 & & 9.7 & 4.3 & 4.3 & 2.0 & 0.93 & & 2.0 & g... & \nodata & 9.5 & 1.22 & & 21411514$+$5806520 \\
106 & 214115.64$+$581933.5 & & 325.315208 & 58.325980 & 5.81 & & 16.9 & 4.8 & 1.1 & 10.2 & 0.90 & & 3.2 & .... & a & 25.3 & 2.34 & & \nodata \\
107 & 214115.87$+$581453.6 & & 325.316126 & 58.248235 & 4.73 & & 32.4 & 6.3 & 0.6 & 1.6 & 0.89 & & 4.8 & .... & a & 25.9 & 1.01 & & 21411586$+$5814537 \\
108 & 214116.28$+$581546.2 & & 325.317836 & 58.262855 & 4.66 & & 11.5 & 4.0 & 0.5 & 2.6 & 0.89 & & 2.5 & .... & c & 26.0 & 1.28 & & \nodata \\
109 & 214117.17$+$581743.4 & & 325.321560 & 58.295408 & 5.08 & & 8.3 & 3.5 & 0.7 & 4.5 & 0.90 & & 2.0 & .... & a & 24.7 & 3.41 & & \nodata \\
110 & 214120.48$+$581641.2 & & 325.335352 & 58.278119 & 5.26 & & 8.9 & 3.7 & 1.1 & 3.2 & 0.90 & & 2.1 & .... & a & 25.5 & 1.73 & & \nodata \\
111 & 214122.22$+$581823.9 & & 325.342617 & 58.306641 & 5.95 & & 466.9 & 22.1 & 1.1 & 162.2 & 0.90 & & 20.6 & g... & \nodata & 20.9 & 1.61 & & 21412220$+$5818238 \\
112 & 214123.52$+$581536.6 & & 325.348041 & 58.260183 & 5.62 & & 42.0 & 7.1 & 1.0 & 7.3 & 0.90 & & 5.5 & .... & a & 25.0 & 1.42 & & 21412352$+$5815371 \\
113 & 214132.38$+$581303.5 & & 325.384949 & 58.217662 & 7.38 & & 27.3 & 6.1 & 3.7 & 16.2 & 0.90 & & 4.1 & .... & a & 22.8 & 2.47 & & \nodata \\
114 & 214135.79$+$581920.5 & & 325.399130 & 58.322373 & 7.97 & & 5.4 & 3.6 & 3.6 & 0.7 & 0.89 & & 1.3 & .... & a & 22.5 & 1.64 & & 21413560$+$5819193 \\
115 & 214135.81$+$581826.3 & & 325.399239 & 58.307314 & 7.63 & & 9.8 & 4.1 & 3.2 & 4.9 & 0.90 & & 2.1 & .... & a & 22.8 & 2.15 & & \nodata \\
116 & 214138.03$+$581759.5 & & 325.408471 & 58.299869 & 7.78 & & 14.8 & 4.8 & 3.2 & 1.8 & 0.89 & & 2.8 & .... & a & 22.6 & 1.25 & & 21413812$+$5817599 \\
117 & 214148.52$+$581915.6 & & 325.452181 & 58.321006 & 9.48 & & 10.5 & 4.8 & 7.5 & 4.8 & 0.89 & & 1.9 & .... & a & 20.5 & 1.89 & & \nodata \\
\enddata

\tablecomments{Column 1: X-ray source number. Column 2: IAU
designation. Columns 3-4: Right ascension and declination for
epoch J2000.0 in degrees. Column 5: Off-axis angle. Columns 6-7:
Estimated net counts from extracted area in total energy band
$(0.5-8.0)$ keV and their 1\/$\sigma$ errors. Column 8: Estimated
background counts from extracted area in total energy band. Column
9: Estimated net counts from extracted area in hard energy band
$(2-8)$ keV. Column 10: Fraction of the PSF at the fiducial energy
of 1.497 keV enclosed within the extracted area. Note that a
reduced PSF fraction (significantly below 90\%) may indicate that
the source is in a crowded region or on the edge of the field.
Column 11: Source significance in sigma. Column 12. Source
anomalies described by a set of flags, see the text
note$^{\rm{a}}$ below for details. Column 13: Variability
characterization based on K-S statistics, see the text
note$^{\rm{b}}$ below for details. Column 14: Source effective
exposure in ks. Column 15: The background-corrected median photon
energy in the total $0.5-8$ keV energy band. Column 16: 2MASS
counterpart.}

\tablenotetext{a}{ Source anomalies:  g = fractional time that source
was on a detector (FRACEXPO from {\em mkarf}) is $<0.9$ ; e = source
on field edge; p = source piled up; s = source on readout streak.}

\tablenotetext{b}{ Source variability:  a = no evidence for
variability; b = possibly variable; c = definitely variable. No test
is performed for sources with fewer than 4 total full-band counts.
No value is reported for sources in chip gaps or on field edges.}

\end{deluxetable}
\clearpage

\begin{deluxetable}{rcrccrrrccrrrrrrc}
\centering \rotate \tabletypesize{\tiny} \tablewidth{0pt}
\tablecolumns{17}

\tablecaption{Infrared Counterparts to X-ray Globule Sources
\label{tbl_nirmir}}

\tablehead{ \multicolumn{2}{c}{Source} &
  &
\multicolumn{8}{c}{NIR} &
  &
\multicolumn{5}{c}{MIR} \\
\cline{1-2} \cline{4-11} \cline{13-17}

\colhead{Seq} & \colhead{CXOU~J} &
  &
\colhead{2MASS} & \colhead{Offset} & \colhead{$J$} & \colhead{$H$} &
\colhead{$K_s$} & \colhead{Flag} & \colhead{Nisini} & \colhead{$K$} &
  &
\colhead{Offset} & \colhead{[3.6]} & \colhead{[4.5]} &
\colhead{[5.8]} & \colhead{Class} \\

\colhead{\#} & \colhead{} &
  &
\colhead{} & \colhead{$\arcsec$} & \colhead{mag} & \colhead{mag} &
\colhead{mag} & \colhead{} & \colhead{\#} & \colhead{mag} &
  &
\colhead{$\arcsec$} & \colhead{mag} & \colhead{mag} &
\colhead{mag} & \colhead{} \\

\colhead{(1)} & \colhead{(2)} &
  &
\colhead{(3)} & \colhead{(4)} & \colhead{(5)} & \colhead{(6)} &
\colhead{(7)} & \colhead{(8)} & \colhead{(9)} & \colhead{(10)} &
  &
\colhead{(11)} & \colhead{(12)} & \colhead{(13)} & \colhead{(14)}
& \colhead{(15)} }

\startdata
41 & 214027.31+581421.1 & & 21402732+5814212 & 0.07 & 14.30 & 13.30 & 12.88 & AAA000 & \nodata & \nodata & & 0.53 & 11.62$\pm 0.01$ & 11.09$\pm 0.01$ & 10.74$\pm 0.02$ & II \\
49 & 214031.58+581755.2 & & 21403159+5817551 & 0.13 & 14.03 & 12.89 & 12.39 & AAA000 & \nodata & \nodata & & 0.26 & 11.51$\pm 0.02$ & 11.16$\pm 0.02$ & 10.23$\pm 0.04$ & II \\
53 & 214036.57+581345.8 & & 21403655+5813458 & 0.13 & 13.51 & 12.58 & 12.24 & AAA000 & \nodata & \nodata & & 0.67 & 12.13$\pm 0.02$ & 11.94$\pm 0.02$ & 12.01$\pm 0.03$ & III \\
55 & 214036.90+581437.9 & & 21403691+5814378 & 0.14 & 11.90 & 10.89 & 10.23 & AAA000 & \nodata & \nodata & & 0.56 & 9.38$\pm 0.00$ & 9.10$\pm 0.01$ & 8.76$\pm 0.01$ & II \\
60 & 214039.62+581609.3 &  & \nodata & \nodata & \nodata & \nodata & \nodata & \nodata & 2 & 15.14 & & 0.78 & 11.29$\pm 0.02$ & 9.42$\pm 0.01$ & 8.31$\pm 0.01$ & I\tablenotemark{a} \\
61 & 214039.87+581834.8 & & 21403988+5818349 & 0.13 & $>18.29$ & 15.30 & 13.36 & UBA000 & 3 & 13.19 & & 0.77 & 11.63$\pm 0.02$ & 10.95$\pm 0.01$ & 10.21$\pm 0.02$ & II \\
62 & 214041.12+581359.0 & & 21404110+5813588 & 0.20 & 12.96 & 12.08 & 11.77 & AAA000 & \nodata & \nodata & & 0.51 & 11.61$\pm 0.01$ & 11.72$\pm 0.02$ & 11.49$\pm 0.02$ & III \\
63 & 214041.16+581511.2 &  & 21404116+5815112 & 0.06 & 12.97 & 11.61 & 10.68 & AAA000 & \nodata & \nodata & & 0.42 & 9.15$\pm 0.01$ & 8.61$\pm 0.00$ & 8.05$\pm 0.01$ & II \\
65 & 214041.56+581425.5 & & 21404156+5814255 & 0.05 & 13.65 & 12.62 & 12.17 & AAA000 & \nodata & \nodata & & 0.69 & 11.40$\pm 0.01$ & 11.20$\pm 0.02$ & 10.68$\pm 0.02$ & II \\
66 & 214041.81+581612.3 &  & \nodata & \nodata & \nodata & \nodata & \nodata & \nodata & \nodata & \nodata & & 1.11 & 11.30$\pm 0.03$ & 8.90$\pm 0.01$ & 7.50$\pm 0.01$ & 0/I\tablenotemark{b} \\
67 & 214041.91+581523.1 &  & 21404191+5815230 & 0.15 & 15.68 & 14.30 & 13.65 & AAA000 & \nodata & \nodata & & 0.19 & 12.69$\pm 0.05$ & 12.49$\pm 0.04$ & $>9.82$ & II \\
68 & 214042.89+581601.0 &  & \nodata & \nodata & \nodata & \nodata & \nodata & \nodata & \nodata & \nodata & & 0.63 & 10.60$\pm 0.01$ & 8.78$\pm 0.01$ & 7.60$\pm 0.01$ & I\tablenotemark{c} \\
70 & 214043.47+581559.7 &  & \nodata & \nodata & \nodata & \nodata & \nodata & \nodata & \nodata & \nodata & & 1.01 & 12.89$\pm 0.07$ & 11.95$\pm 0.20$ & $>9.85$ & I/II \\
71 & 214043.64+581618.9 &  & 21404365+5816190 & 0.11 & $>$17.89 & $>$16.09 & 13.51 & UUA000 & 10 & 13.36 & & 0.67 & 9.97$\pm 0.01$ & 8.72$\pm 0.01$ & 8.00$\pm 0.01$ & I \\
72 & 214044.34+581513.3 & & 21404433+5815133 & 0.09 & 16.05 & 14.59 & 13.60 & AAA000 & \nodata & \nodata & & 0.56 & 12.42$\pm 0.03$ & 11.64$\pm 0.02$ & 10.60$\pm 0.05$ & II \\
73 & 214044.84+581605.1 &  & 21404485+5816050 & 0.19 & $>$15.85 & 14.28 & 12.89 & UAA000 & 11 & 12.77 & & 0.27 & 11.73$\pm 0.02$ & 11.15$\pm 0.02$ & 10.68$\pm 0.03$ & II \\
74 & 214044.85+581503.4 & & 21404485+5815033 & 0.13 & 14.62 & 13.35 & 12.66 & AAA000 & \nodata & \nodata & & 0.56 & 12.29$\pm 0.02$ & 11.40$\pm 0.02$ & 10.96$\pm 0.04$ & II \\
76 & 214045.18+581559.8 &  & \nodata & \nodata & \nodata & \nodata & \nodata & \nodata & \nodata & \nodata & & 0.71 & 11.95$\pm 0.02$ & 10.82$\pm 0.02$ & 10.03$\pm 0.02$ & I \\
77 & 214045.51+581511.4 & & 21404551+5815115 & 0.16 & 14.65 & 13.71 & 13.11 & AAA000 & \nodata & \nodata & & 0.58 & 12.54$\pm 0.03$ & 12.21$\pm 0.03$ & 11.82$\pm 0.10$ & II \\
78 & 214045.53+581602.9 &  & 21404553+5816027 & 0.19 & 15.68 & 13.73 & 12.85 & AAA000 & \nodata & \nodata & & 1.09 & 12.23$\pm 0.02$ & 11.84$\pm 0.03$ & $>10.37$ & II \\
80 & 214045.79+581549.0 &  & \nodata & \nodata & \nodata & \nodata & \nodata & \nodata & \nodata & \nodata & & 0.51 & 12.59$\pm 0.03$ & 11.19$\pm 0.02$ & 10.04$\pm 0.02$ & I \\
81 & 214046.49+581523.2 & & 21404649+5815230 & 0.21 & 12.81 & 11.95 & 11.65 & AAA000 & \nodata & \nodata & & 0.81 & 11.37$\pm 0.02$ & 11.33$\pm 0.02$ & 11.39$\pm 0.19$ & III \\
82 & 214046.89+581533.3 & & 21404688+5815334 & 0.07 & 15.30 & 13.55 & 12.63 & AAA000 & \nodata & \nodata & & 0.53 & 11.97$\pm 0.03$ & 11.80$\pm 0.02$ & 11.15$\pm 0.17$ & II \\
85 & 214048.03+581537.9 & & 21404803+5815376 & 0.25 & 13.89 & 12.95 & 12.67 & AAA000 & \nodata & \nodata & & 0.78 & 12.08$\pm 0.02$ & 11.91$\pm 0.02$ & 10.77$\pm 0.14$ & II \\
87 & 214049.09+581709.3 &  & 21404908+5817093 & 0.04 & 14.14 & 12.86 & 12.13 & AAA000 & \nodata & \nodata & & 0.42 & 10.77$\pm 0.01$ & 10.19$\pm 0.01$ & 9.69$\pm 0.01$ & II \\
\enddata

\tablecomments{Column 1: X-ray source number. Column 2: IAU
designation. Columns 3: 2MASS name. Column 4: $Chandra-$2MASS
positional offset. Columns 5-7: 2MASS $JHK_s$ magnitudes. Column
8: 2MASS photometry quality and confusion-contamination flag.
Column 9-10: NIR identifier and $K-$band magnitude from
\citet{Nisini01}. Column 11: $Chandra-Spitzer$ positional offset.
Columns 12-14: $Spitzer$ [3.6], [4.5], and [5.8] magnitudes and
their formal statistical errors. Column 15: Suggested PMS class
based on a presence of NIR counterpart and positions on
color-color NIR-MIR diagrams (Figure \ref{nir_mir_fig}).}

\tablenotetext{a}{The X-ray source \#60 is offset by 3.2$\arcsec$
to the north of the radio continuum source VLA 1, and by
5.0$\arcsec$ to the northwest of the millimeter source BIMA 1 of
\citet{Beltran02}.}

\tablenotetext{b}{The  X-ray source \#66 is within 0.5$\arcsec$ of
the radio continuum source VLA 2 and millimeter source BIMA 2 of
\citet{Beltran02}. This is also millimeter source A of
\citet{Codella01}.}

\tablenotetext{c}{The X-ray source \#68 is within 1.0$\arcsec$ of
the radio continuum source VLA 3 and millimeter source BIMA 3 of
\citet{Beltran02}. This is also millimeter source B of
\citet{Codella01}.}

\end{deluxetable}

\clearpage

\begin{deluxetable}{ccccccccrrrrcrr}
\centering \rotate \tabletypesize{\tiny} \tablewidth{0pt}
\tablecolumns{15}

\tablecaption{NIR Globule Sample of \citet{Nisini01}
\label{tbl_nisini}}

\tablehead{ \multicolumn{4}{c}{NIR Nisini01} & &
\multicolumn{1}{c}{$Chandra$} & & \multicolumn{5}{c}{$2MASS$} & & \multicolumn{2}{c}{$Spitzer$} \\
\cline{1-4} \cline{6-6} \cline{8-12} \cline{14-15}

\colhead{Seq} & \colhead{$\alpha_{\rm J2000}$} &
\colhead{$\delta_{\rm J2000}$} & \colhead{$K$} & & \colhead{Seq} &
& \colhead{$2MASS$} & \colhead{$J$} & \colhead{$H$} &
\colhead{$K_s$} & \colhead{Flag} & &
\colhead{[3.6]} & \colhead{[4.5]} \\

\colhead{\#NIR} & \colhead{deg} & \colhead{deg} & \colhead{mag} &
& \colhead{\#} & & \colhead{} & \colhead{mag} & \colhead{mag} &
\colhead{mag} & \colhead{} & &
\colhead{mag} & \colhead{mag} \\

\colhead{(1)} & \colhead{(2)} & \colhead{(3)} & \colhead{(4)} & &
\colhead{(5)} & & \colhead{(6)} & \colhead{(7)} & \colhead{(8)} &
\colhead{(9)} & \colhead{(10)} & &
\colhead{(11)} & \colhead{(12)} \\
}

\startdata
1 & 325.162790 & 58.293791 & 15.81 & & \nodata & & \nodata & \nodata & \nodata & \nodata & \nodata & & 14.62$\pm0.09$ & 14.63$\pm0.08$\\
2 & 325.165100 & 58.269072 & 15.14 & & 60 & & \nodata & \nodata & \nodata & \nodata & \nodata & & 11.29$\pm0.02$ & 9.42$\pm0.01$\\
3 & 325.166183 & 58.309711 & 13.19 & & 61 & & 21403988+5818349 & $>$18.29 & 15.30 & 13.36 & UBA000 & & 11.63$\pm0.02$ & 10.95$\pm0.01$\\
4 & 325.168090 & 58.279117 & 15.88 & & \nodata & & \nodata & \nodata & \nodata & \nodata & \nodata & & 15.20$\pm0.14$ & 13.99$\pm0.11$\\
5 & 325.168616 & 58.307255 & 13.59 & & \nodata & & 21404046+5818261 & $>$18.28 & 15.64 & 13.85 & UCA000 & & 12.81$\pm0.03$ & 12.55$\pm0.03$\\
6 & 325.170881 & 58.281258 & 13.87 & & \nodata & & 21404101+5816525 & $>$17.93 & 15.34 & 13.93 & UBA000 & & 13.30$\pm0.03$ & 13.20$\pm0.04$\\
7 & 325.170842 & 58.270039 & 13.02 & & \nodata & & 21404100+5816121 & $>$17.55 & $>$15.89 & 13.59 & UUA000 & & 11.11$\pm0.06$ & 9.89$\pm0.03$\\
8 & 325.172657 & 58.277168 & 12.28 & & \nodata & & 21404143+5816378 & $>$18.30 & 15.60 & 12.78 & UBA000 & & 9.56$\pm0.01$ & 8.28$\pm0.01$\\
9 & 325.180710 & 58.287924 & 15.31 & & \nodata & & \nodata & \nodata & \nodata & \nodata & \nodata & & 13.13$\pm0.03$ & 12.51$\pm0.03$\\
10 & 325.181875 & 58.271961 & 13.36 & & 71 & & 21404365+5816190 & $>$17.89 & $>$16.09 & 13.51 & UUA000 & & 9.97$\pm0.01$ & 8.72$\pm0.01$\\
11 & 325.186903 & 58.268063 & 12.77 & & 73 & & 21404485+5816050 & $>$15.85 & 14.28 & 12.89 & UAA000 & & 11.73$\pm0.02$ & 11.15$\pm0.02$\\
12 & 325.190370 & 58.274338 & 14.61 & & \nodata & & 21404737+5816372 & $>$17.07 & 16.13 & 14.63 & UDB000 & & 13.80$\pm0.06$ & 12.99$\pm0.03$\\
13 & 325.190970 & 58.261361 & 15.36 & & \nodata & & \nodata & \nodata & \nodata & \nodata & \nodata & & 14.76$\pm0.30$ & 13.89$\pm0.10$\\
14 & 325.196290 & 58.261592 & 15.04 & & \nodata & & \nodata & \nodata & \nodata & \nodata & \nodata & & 14.14$\pm0.10$ & 13.99$\pm0.10$\\
15 & 325.197120 & 58.283615 & 15.28 & & \nodata & & \nodata & \nodata & \nodata & \nodata & \nodata & & 14.13$\pm0.05$ & 14.13$\pm0.06$\\
16 & 325.197381 & 58.277023 & 14.50 & & \nodata & & 21404737+5816372 & $>$17.08 & 16.13 & 14.63 & UDB000 & & 13.67$\pm0.04$ & 13.32$\pm0.04$\\
17 & 325.202541 & 58.266933 & 11.72 & & \nodata & & 21404860+5816009 & 15.88 & 13.45 & 11.88 & AAA000 & & 10.09$\pm0.01$ & 9.35$\pm0.01$\\
18 & 325.230180 & 58.275707 & 15.81 & & \nodata & & \nodata & \nodata & \nodata & \nodata & \nodata & & 15.10$\pm0.09$ & 15.26$\pm0.17$\\
19 & 325.238759 & 58.278313 & 14.38 & & \nodata & & 21405730+5816419 & $>$16.64 & $>$15.41 & 14.55 & UUB000 & & 13.64$\pm0.07$ & 12.65$\pm0.04$\\
20 & 325.116689 & 58.277267 & 14.92 & & \nodata & & 21402800+5816381 & 15.89 & 15.03 & $>$12.35 & BCUcc0 & & 12.32$\pm0.02$ & 12.52$\pm0.03$\\
\enddata

\tablecomments{Column 1: \citet{Nisini01} NIR source number.
Columns 2-3: Right ascension and declination for epoch J2000.0 in
degrees. Column 4: $K-$band magnitude from \citet{Nisini01}.
Column 5: $Chandra$ source number. Column 6: $2MASS$ source.
Columns 7-9: $2MASS$ $JHK_s$ magnitudes. Column 10: $2MASS$
photometry quality and confusion and contamination flag. Columns
11-12: $Spitzer$ [3.6], [4.5] magnitudes and their formal
statistical errors.}
\end{deluxetable}

\clearpage

\thispagestyle{empty}
\begin{deluxetable}{rccrrrrrccccccccccc}
\centering \rotate \tabletypesize{\tiny} \tablewidth{0pt}
\tablecolumns{19}

\tablecaption{X-ray Properties of Globule Sources
\label{tbl_xray}}

\tablehead{ \multicolumn{2}{c}{Source} & &
\multicolumn{5}{c}{X-ray Photometry} & & \multicolumn{10}{c}{X-ray Spectra and Luminosities} \\
\cline{1-2} \cline{4-8} \cline{10-19}

\colhead{Seq} & \colhead{CXOU~J} & & \colhead{NetFull} &
\colhead{$\Delta$ NetFull} & \colhead{EffExp} & \colhead{P$_{KS}$}
& \colhead{MedE}\tablenotemark{b} & & \colhead{$\log N_H$} &
\colhead{$kT$} & \colhead{$\log L_t$} & \colhead{$\Delta \log
L_t$} & \colhead{$\log L_h$} & \colhead{$\Delta \log L_h$} &
\colhead{$\log L_{t,c}$} & \colhead{$\Delta \log L_{t,c}$} & \colhead{$\log L_{h,c}$} & \colhead{$\Delta \log L_{h,c}$} \\

\colhead{\#} & \colhead{} & & \colhead{cnts} & \colhead{cnts} &
\colhead{ks} & \colhead{} & \colhead{keV} & & \colhead{cm$^{-2}$}
& \colhead{keV} & \colhead{erg s$^{-1}$} & \colhead{erg s$^{-1}$}
& \colhead{erg s$^{-1}$} & \colhead{erg s$^{-1}$} & \colhead{erg
s$^{-1}$} & \colhead{erg s$^{-1}$} & \colhead{erg s$^{-1}$} & \colhead{erg s$^{-1}$} \\

\colhead{(1)} & \colhead{(2)} & & \colhead{(3)} & \colhead{(4)} &
\colhead{(5)} & \colhead{(6)} & \colhead{(7)} & & \colhead{(8)} &
\colhead{(9)} & \colhead{(10)} & \colhead{(11)} & \colhead{(12)} &
\colhead{(13)} & \colhead{(14)} & \colhead{(15)} & \colhead{(16)}
& \colhead{(17)} }

\startdata
41 & 214027.31+581421.1 & & 10.8 & 3.8 & 28.4 & 0.895 & 1.47 & & {\it 21.61} & 1.74 & 29.16 & 0.19 & 28.86 & 0.34 & 29.39 & 0.28 & 28.89 & 0.36 \\
49 & 214031.58+581755.2 & & 26.8 & 5.7 & 29.3 & 0.207 & 1.61 & & {\it 21.74} & 1.67 & 29.51 & 0.13 & 29.24 & 0.20 & 29.80 & 0.20 & 29.28 & 0.20\\
53 & 214036.57+581345.8 & & 28.8 & 5.9 & 28.7 & 0.440 & 1.45 & & {\it 21.31} & 2.24 & 29.41 & 0.12 & 29.12 & 0.19 & 29.53 & 0.15 & 29.13 & 0.19\\
55 & 214036.90+581437.9 & & 10.8 & 3.8 & 29.4 & 0.076 & 1.41 & & {\it 21.84} & 1.00 & 29.19 & 0.19 & 28.71 & 0.35 & 29.71 & 0.29 & 28.77 & 0.35\\
60 & 214039.62+581609.3 & & 26.9 & 5.7 & 29.5 & 0.728 & 4.62 & & 22.87 & 4.60 & 30.01 & 0.13 & 30.01 & 0.20 & 30.52 & 0.58 & 30.27 & 0.29\\
61 & 214039.87+581834.8 & & 113.9 & 11.2 & 26.6 & 0.000 & 3.54 & & 22.41 & $>15$ & 30.62 & 0.07 & 30.60 & 0.09 & 30.86 & 0.24 & 30.69 & 0.11\\
62 & 214041.12+581359.0 & & 71.8 & 9.0 & 27.4 & 0.362 & 1.45 & & 21.33 & 2.16 & 30.12 & 0.09 & 29.82 & 0.14 & 30.25 & 0.12 & 29.84 & 0.14\\
63 & 214041.16+581511.2 & & 46.9 & 7.4 & 29.4 & 0.854 & 2.42 & & 22.38 & 1.95 & 30.18 & 0.08 & 30.11 & 0.10 & 30.72 & 0.32 & 30.27 & 0.21\\
65 & 214041.56+581425.5 & & 8.9 & 3.5 & 21.6 & 0.716 & 1.76 & & {\it 21.54} & 4.33 & 29.36 & 0.21 & 29.22 & 0.42 & 29.49 & 0.37 & 29.23 & 0.42\\
66 & 214041.81+581612.3 & & 7.9 & 3.4 & 29.4 & 0.122 & 5.99 & & 23.98 & 2.39 & 29.84 & 0.29 & 29.84 & 0.30 & 31.69 & 0.72 & 31.31 & 0.43\\
67 & 214041.91+581523.1 & & 14.9 & 4.4 & 26.7 & 0.629 & 2.28 & & {\it 22.05} & 2.34 & 29.33 & 0.15 & 29.21 & 0.16 & 29.66 & 0.44 & 29.27 & 0.24\\
68 & 214042.89+581601.0 & & 4.9 & 2.8 & 29.3 & 0.567 & 5.61 & & 23.93 & 2.46 & 29.61 & 0.33 & 29.61 & 0.34 & 31.35 & 0.81 & 30.98 & 0.48\\
70 & 214043.47+581559.7 & & 25.9 & 5.6 & 29.3 & 0.003 & 3.28 & & 22.80 & 2.32 & 29.89 & 0.16 & 29.88 & 0.17 & 30.58 & 0.53 & 30.19 & 0.31\\
71 & 214043.64+581618.9 & & 8.9 & 3.5 & 29.3 & 0.885 & 3.85 & & 22.69 & 2.50 & 29.53 & 0.27 & 29.51 & 0.28 & 30.13 & 0.68 & 29.76 & 0.40\\
72 & 214044.34+581513.3 & & 3.9 & 2.5 & 21.8 & 0.849 & 1.39 & & {\it 22.10} & 0.84 & 28.85 & 0.28 & 28.45 & 0.57 & 29.70 & 0.41 & 28.57 & 0.57\\
73 & 214044.84+581605.1 & & 15.9 & 4.5 & 29.3 & 0.266 & 2.20 & & 22.28 & 1.93 & 29.54 & 0.16 & 29.45 & 0.18 & 30.03 & 0.45 & 29.58 & 0.25\\
74 & 214044.85+581503.4 & & 39.9 & 6.8 & 27.5 & 0.336 & 1.61 & & {\it 21.90} & 1.95 & 29.89 & 0.11 & 29.70 & 0.16 & 30.21 & 0.18 & 29.76 & 0.17\\
76 & 214045.18+581559.8 & & 52.9 & 7.8 & 29.2 & 0.001 & 4.23 & & 23.19 & 2.55 & 30.46 & 0.11 & 30.46 & 0.16 & 31.36 & 0.54 & 30.99 & 0.27\\
77 & 214045.51+581511.4 & & 6.9 & 3.2 & 22.9 & 0.795 & 1.74 & & {\it 21.56} & 0.55 & 28.82 & 0.24 & 27.62 & 0.48 & 29.30 & 0.42 & 27.66 & 0.48\\
78 & 214045.53+581602.9 & & 9.9 & 3.7 & 29.2 & 0.875 & 1.91 & & {\it 22.28} & 1.19 & 28.92 & 0.27 & 28.75 & 0.58 & 29.67 & 0.39 & 28.91 & 0.58\\
80 & 214045.79+581549.0 & & 10.9 & 3.8 & 29.2 & 0.547 & 4.33 & & 22.69 & 2.26 & 29.34 & 0.19 & 29.33 & 0.29 & 29.99 & 0.72 & 29.59 & 0.36\\
81 & 214046.49+581523.2 & & 11.9 & 4.0 & 11.9 & \nodata\tablenotemark{a} & 1.57 & & {\it 20.74} & 5.70 & 29.69 & 0.18 & 29.49 & 0.38 & 29.72 & 0.26 & 29.50 & 0.38\\
82 & 214046.89+581533.3 & & 3.9 & 2.5 & 22.0 & 0.337 & 1.42 & & {\it 22.19} & 0.41 & 28.80 & 0.28 & 28.08 & 0.62 & 30.29 & 0.45 & 28.26 & 0.63\\
85 & 214048.03+581537.9 & & 25.9 & 5.6 & 29.0 & 0.494 & 1.34 & & {\it 21.43} & 0.76 & 29.18 & 0.12 & 28.23 & 0.19 & 29.49 & 0.21 & 28.26 & 0.19\\
87 & 214049.09+581709.3 & & 17.9 & 4.8 & 28.9 & 0.672 & 1.93 & & {\it 21.87} & 0.99 & 29.03 & 0.14 & 28.56 & 0.25 & 29.57 & 0.31 & 28.63 & 0.25\\
\enddata

\tablecomments{Column 1: X-ray source number. Column 2: IAU
designation. Columns 3-4: Estimated net counts from extracted area
in total energy band (0.5-8.0) keV and their 1$\sigma$ errors.
Column 5: Source's effective exposure. Column 6: Probability of
accepting the null hypothesis of a constant source from the
nonparametric one-sample Kolmogorov-Smirnov test. Column 7: The
background-corrected median photon energy in the total energy
band. Column 8: Estimated column density. Italic font indicates
that a column density parameter was fixed at the value
corresponding to a visual absorption, derived from NIR
color-magnitude diagram, and the relationship $N_H = 2.0 \times
10^{21} A_V$ \citep{Ryter96} was applied. Column 9: Estimated
plasma temperature. Columns 10-17: Observed and corrected for
absorption X-ray luminosities, obtained from our spectral
analysis, and their errors estimated from simulations. $h=$ hard
band $(2.0-8.0)$ keV, $t=$ total band $(0.5-8.0)$ keV, $c=$
corrected for absorption.}

\tablenotetext{a}{Source \#81 lies on a chip gap.}

\tablenotetext{b}{For plotting purposes only, some quantities are
reported to two decimal places.}

\end{deluxetable}

\clearpage

\begin{deluxetable}{lrccccccccc}
\centering \tabletypesize{\tiny} \tablewidth{0pt}
\tablecolumns{11}

\tablecaption{X-rays from Protostars \label{tbl_xrayprotostars}}

\tablehead{ \multicolumn{6}{c}{IR/Submm} & &
\multicolumn{4}{c}{X-ray} \\
\cline{1-6} \cline{8-11}

\colhead{Name} & \colhead{$D$} & \colhead{$L_{{\rm bol}}$} &
\colhead{$M_{{\rm env}}$} & \colhead{Class} & \colhead{Ref} & &
\colhead{Obs} & \colhead{Exp} & \colhead{Det?} & \colhead{Ref} \\

\colhead{}  & \colhead{(pc)} & \colhead{($L_{\odot}$)} &
\colhead{($M_{\odot}$)} & \colhead{} & \colhead{} & & \colhead{} &
\colhead{ks} & \colhead{} & \colhead{} \\

\colhead{(1)} & \colhead{(2)} & \colhead{(3)} & \colhead{(4)} &
\colhead{(5)} & \colhead{(6)} & & \colhead{(7)} & \colhead{8)} &
\colhead{(9)} & \colhead{(10)}}

\startdata
\multicolumn{11}{c}{\bf (a) Very Young Protostars Observed in X-rays} \\
L1448~NW        & 300 & $<2.8$    & 1.2   & 0   & 1 & & CXO &  68 & N & 13 \\
IRS~3B          & 300 & $9.3-9.6$ & 0.5   & 0/1 & 2 & & CXO &  68 & N & 13 \\
IRS~3A          & 300 & $1.3-1.6$ & 0.15  & 1   & 2 & & CXO &  68 & Y?& 13 \\
L1448~C         & 300 & 8.3       & 1.1   & 0   & 1 & & CXO &  68 & N & 13 \\
IRAS~03256+3055 & 320 & 1         & 0.34  & 0   & 1 & & XMM & 235 & N & 14 \\
NGC~1333~I2     & 320 & 43        & 1.5   & 0   & 1 & & CXO &  38 & N & 15 \\
                &     &           &       &     &   & & XMM & 235 & N & 14 \\
NGC~1333~I4~A   & 320 & 18        & 5.8   & 0   & 1 & & CXO &  38 & N & 15 \\
                &     &           &       &     &   & & XMM & 235 & N & 14 \\
NGC~1333~I4~B   & 320 & 17        & 3.1   & 0   & 1 & & CXO &  38 & N & 15 \\
                &     &           &       &     &   & & XMM & 235 & N & 14 \\
IC~348~HH211-MM & 310 & 3.6       & 0.8   & 0   & 1 & & XMM & 207 & N & 16 \\
IC~348~MMS      & 310 & 8         & 1.1   & 0   & 3 & & XMM & 207 & N & 16 \\
IRAM~04191      & 140 & 0.12      & 0.48  & 0   & 1 & & CXO &  20 & N?& 17 \\
L~1527          & 140 & 1.9       & 0.8   & 0/1 & 1 & & CXO &  20 & N?& 17 \\
OMC~3-MM~6      & 450 & $<100$    & 6.1   & 0   & 1 & & CXO &  89 & N & 18 \\
OMC~3-MM~9      & 450 & $130$     & 1.9   & 1   & 1 & & CXO &  89 & N & 18 \\
L~1641-VLA~1    & 460 & $<50$     & 1.6   & 0?  & 1 & & CXO &  20 & N & 19 \\
NGC~2024-FIR~5  & 450 & $\ga 10$  & 15    & 0   & 4 & & CXO &  77 & N & 20 \\
NGC~2024-FIR~6  & 450 & $\ga 15$  & 6     & 0   & 4 & & CXO &  77 & N & 20 \\
HH~25~MMS       & 450 & 7.2       & 1.2   & 0   & 1 & & CXO &  63 & N & 21 \\
HH~24~MMS       & 450 & $<18$     & 2.7   & 0   & 1 & & CXO &  63 & N & 21 \\
LBS17-H         & 450 & 1.5       & 2.8   & 0   & 1 & & CXO &  63 & N & 21 \\
VLA~1623        & 160 & $<2$      & 0.8   & 0   & 1 & & CXO &  96 & N & 22 \\
IRAS~16293-2422 & 160 & 21        & 4.6   & 0   & 5 & & CXO &  30 & N?& 17 \\
Trifid-TC3      &1680 & 1000      & 60    & 0?  & 4 & & CXO &  58 & N & 23 \\
Trifid-TC4      &1680 & 520-2400  & 60    & 0?  & 6 & & CXO &  58 & Y?& 23 \\
Serp-S68~N      & 310 & 4.4       & 1.1   & 0   & 5 & & XMM & 250 & N & 24 \\
Serp-FIRS~1     & 310 & 45        & 3.6   & 0/1 & 5 & & XMM & 250 & N & 24 \\
Serp-SMM~5      & 310 & 3.6       & 1.4   & 1   & 5 & & XMM & 250 & N & 24 \\
Serp-SMM~4      & 310 & 5.2       & 1.5   & 0   & 5 & & XMM & 250 & N & 24 \\
Serp-SMM~3      & 310 & 4.9       & 1     & 0   & 5 & & XMM & 250 & N & 24 \\
Serp-SMM~2      & 310 & 4         & 0.36  & 0   & 5 & & XMM & 250 & N & 24 \\
IRS~7B          & 170 & $<2.3$    & 0.2   & 0   & 28,7 & & XMM & 195 & Y & 25 \\
                &     &           &       &     &   & & CXO & 77\tablenotemark{a} & Y & 25 \\
IC~1396N-BIMA~2 & 750 & 150       & 5.1   & 1/0 & 8 & & CXO &  30 & Y & 26 \\
&&&&&&&&&&\\
\multicolumn{11}{c}{\bf (b) Ophiuchus and Serpens Protostars Observed in X-rays} \\
CRBR~12         & 160 & \nodata   &\nodata   & 1 & 9 & & CXO &  96 & N & 22 \\
GSS~30~IRS-1    & 160 & 26        & 0.06     & 1 & 10 & & CXO &  96 & N & 22 \\
GSS~30~IRS-3    & 160 & 0.5       & 0.11     & 1 & 11 & & CXO &  96 & N & 22 \\
GY~91           & 160 & \nodata   &\nodata   & 1 & 9 & & CXO &  96 & Y & 22 \\
WL~12           & 160 & 4.1       & 0.03     & 1 & 10 & & CXO & 100 & Y & 27 \\
LFAM~26         & 160 & \nodata   &\nodata   & 1 & 9 & & CXO & 100 & N & 27 \\
EL~29           & 160 & 41        & 0.09     & 1 & 10 & & CXO & 100 & Y & 27 \\
WL~6            & 160 & 2         &$\la 0.02$& 1 & 11 & & CXO & 100 & Y & 27 \\
CRBR~85         & 160 & \nodata   & \nodata  & 1 & 9 & & CXO & 100 & N & 27 \\
IRS~43          & 160 & 10.1      & 0.05     & 1 & 10 & & CXO & 100 & Y & 27 \\
IRS~44          & 160 & 13        & 0.05     & 1 & 10 & & CXO & 100 & Y & 27 \\
IRS~46          & 160 & 0.6       & 0.02     & 1 & 11 & & CXO & 100 & Y & 27 \\
IRS~48          & 160 & 11.1      & 0.06     & 1 & 11 & & CXO & 100 & N & 27 \\
IRS~51          & 160 & 1.3       & 0.06     & 1 & 9 & & CXO & 100 & Y & 27 \\
IRS~54          & 160 & 11.0      & $<0.03$  & 1 & 10 & & CXO & 100 & N & 27 \\
ISO~241         & 310 & \nodata   & \nodata  & 1 & 12 & & XMM & 250 & N & 24 \\
ISO~249         & 310 & \nodata   & \nodata  & 1 & 12 & & XMM & 250 & N & 24 \\
ISO~250         & 310 & \nodata   & \nodata  & 1 & 12 & & XMM & 250 & N & 24 \\
ISO~253         & 310 & \nodata   & \nodata  & 1 & 12 & & XMM & 250 & N & 24 \\
ISO~254         & 310 & \nodata   & \nodata  & 1 & 12 & & XMM & 250 & N & 24 \\
ISO~258a        & 310 & \nodata   & \nodata  & 1 & 12 & & XMM & 250 & N & 24 \\
ISO~260         & 310 & \nodata   & \nodata  & 1 & 12 & & XMM & 250 & N & 24 \\
ISO~265         & 310 & \nodata   & \nodata  & 1 & 12 & & XMM & 250 & Y & 24 \\
ISO~270         & 310 & \nodata   & \nodata  & 1 & 12 & & XMM & 250 & N & 24 \\
ISO~276         & 310 & \nodata   & \nodata  & 1 & 12 & & XMM & 250 & N & 24 \\
ISO~277         & 310 & \nodata   & \nodata  & 1 & 12 & & XMM & 250 & N & 24 \\
ISO~306         & 310 & \nodata   & \nodata  & 1 & 12 & & XMM & 250 & Y & 24 \\
ISO~308         & 310 & \nodata   & \nodata  & 1 & 12 & & XMM & 250 & N & 24 \\
ISO~312         & 310 & \nodata   & \nodata  & 1 & 12 & & XMM & 250 & Y & 24 \\
ISO~313         & 310 & \nodata   & \nodata  & 1 & 12 & & XMM & 250 & N & 24 \\
ISO~326         & 310 & \nodata   & \nodata  & 1 & 12 & & XMM & 250 & N & 24 \\
ISO~327         & 310 & \nodata   & \nodata  & 1 & 12 & & XMM & 250 & N & 24 \\
ISO~330         & 310 & \nodata   & \nodata  & 1 & 12 & & XMM & 250 & Y & 24 \\
ISO~331         & 310 & \nodata   & \nodata  & 1 & 12 & & XMM & 250 & N & 24 \\
\enddata

\tablecomments{Column 1: Name of protostellar object. Column 2:
Approximate distance.  Column 3: Bolometric luminosity. Column 4:
Envelope mass. Column 5: Evolutionary classification. Column 6:
Reference for columns 2-5.  Column 7: X-ray observatory. Column 8:
X-ray exposure time. Column 9: X-ray detection flag (Y = detected,
N = non-detected).  Column: Reference for columns 7-9.}
\tablecomments{References: 1 = \citet{Froebrich05b}, 2 =
\citet{Ciardi03}, 3 = \citet{Eisloffel03}, 4 = \citet{Andre00}, 5
= \citet{Froebrich05}, 6 = \citet{Lefloch00}, 7 =
\citet{Nutter05}, 8 = \citet{Beltran02}, 9 = \citet{Bontemps01},
10 = \citet{Andre94}, 11 = \citet{Bontemps96}, 12 =
\citet{Kaas04}, 13 = \citet{Tsujimoto05}, 14 =
\citet{Preibisch03}, 15 = \citet{Getman02}, 16 =
\citet{Preibisch04}, 17 = unpublished (image examined by us), 18 =
\citet{Tsujimoto02a}, 19 = \citet{Pravdo01}, 20 =
\citet{Skinner03}, 21 = \citet{Simon04}, 22 = \citet{Gagne04}, 23
= \citet{Rho04}, 24 = \citet{Preibisch04a}, 25 =
\citet{Hamaguchi05}, 26 = This work, 27 = \citet{Imanishi01}, 28 =
\citet{Saraceno96a}}

\tablenotetext{a}{\citet{Hamaguchi05} reports $\sim 32$ X-ray
counts from the source in two $Chandra$ observations of the R CrA
with 57 ks exposure. From visual inspection, we confirm that
another $\sim 50$ counts are seen from the source during the third
$Chandra$ exposure of $20$ ks.}

\end{deluxetable}

\clearpage

\begin{deluxetable}{rllc}
\centering \tabletypesize{\tiny} \tablewidth{0pt}
\tablecolumns{4}

\tablecaption{Infrared properties of X-ray observed Class I
protostars in Ophiuchus and Serpens \label{tbl_classi_irprop}}

\tablehead{ \colhead{Name} & \colhead{$\alpha_{2-14}$} &
\colhead{$K$} & \colhead{Ref.} \\

\colhead{} & \colhead{} & \colhead{mag} & \colhead{} \\

\colhead{(1)} & \colhead{(2)} & \colhead{(3)} & \colhead{(4)} \\
}

\startdata
CRBR~12               & ~0.91 & 12.04 & 1,2 \\
GSS~30~IRS-1          & ~1.2~ & 12.85 & 1,2 \\
GSS~30~IRS-3$=$LFAM~1 & ~1.08 & 13.59 & 1,2 \\
GY~91$=$CRBR~42       & ~0.7~ & 12.51 & 1,2 \\
WL~12                 & ~1.04 & 10.18 & 1,2 \\
LFAM~26$=$GY~197      & ~1.25 & 14.63 & 1,2 \\
EL~29                 & ~0.98 & ~7.54 & 1,2 \\
WL~6                  & ~0.59 & 10.04 & 1,2 \\
CRBR~85               & ~1.48 & 13.21 & 1,2 \\
IRS~43$=$GY~265       & ~0.98 & ~9.46 & 1,2 \\
IRS~44$=$GY~269       & ~1.57 & ~9.65 & 1,2 \\
IRS~46$=$GY~274       & ~0.94 & 11.46 & 1,2 \\
IRS~48$=$GY~304       & ~0.18 & ~7.42 & 1,2 \\
IRS~51$=$GY~315       & -0.04 & ~8.93 & 1,2 \\
IRS~54$=$GY~378       & ~1.76 & 10.87 & 1,2 \\
ISO~241               & ~1.74 & 16.11 & 3 \\
ISO~249$=$EC~37       & ~1.98 & 13.62 & 3 \\
ISO~250$=$DEOS        & ~2.06 & 11.67 & 3 \\
ISO~253$=$EC~40       & ~1.23 & 14.90 & 3 \\
ISO~254$=$EC~38       & ~0.82 & 12.46 & 3 \\
ISO~258a$=$EC~41      & ~0.96 & 14.80 & 3 \\
ISO~260               &\nodata& 16.76 & 3 \\
ISO~265$=$EC~53       & ~0.98 & 11.32 & 3 \\
ISO~270               & ~2.56 & 17.70 & 3 \\
ISO~276$=$GCNM~53     & ~2.02 & 14.80 & 3 \\
ISO~277$=$EC~63       &\nodata& 14.88 & 3 \\
ISO~306$=$EC~80       & ~0.44 & 13.75 & 3 \\
ISO~308$=$HCE~170     &\nodata& 16.62 & 3 \\
ISO~312$=$EC~89       & ~0.85 & 11.24 & 3 \\
ISO~313               &\nodata& 13.46 & 3 \\
ISO~326$=$EC~103      & ~0.62 & 11.56 & 3 \\
ISO~327$=$HCE~175     & ~0.53 & 15.41 & 3 \\
ISO~330$=$HB~1        & ~2.35 & 14.37 & 3 \\
ISO~331               & ~2.51 & 15.37 & 3 \\
\enddata

\tablecomments{Column 1: Name of protostellar object. Column 2: IR
spectral index. Column 3: $K$-band magnitude. Column 4: 1 =
\citet{Bontemps01}, 2 = \citet{Duchene04}, 3 = \citet{Kaas04}}

\end{deluxetable}

\clearpage

\clearpage

\clearpage

\clearpage

\clearpage

\clearpage

\centerline{\includegraphics[angle=0.,width=6.5in]{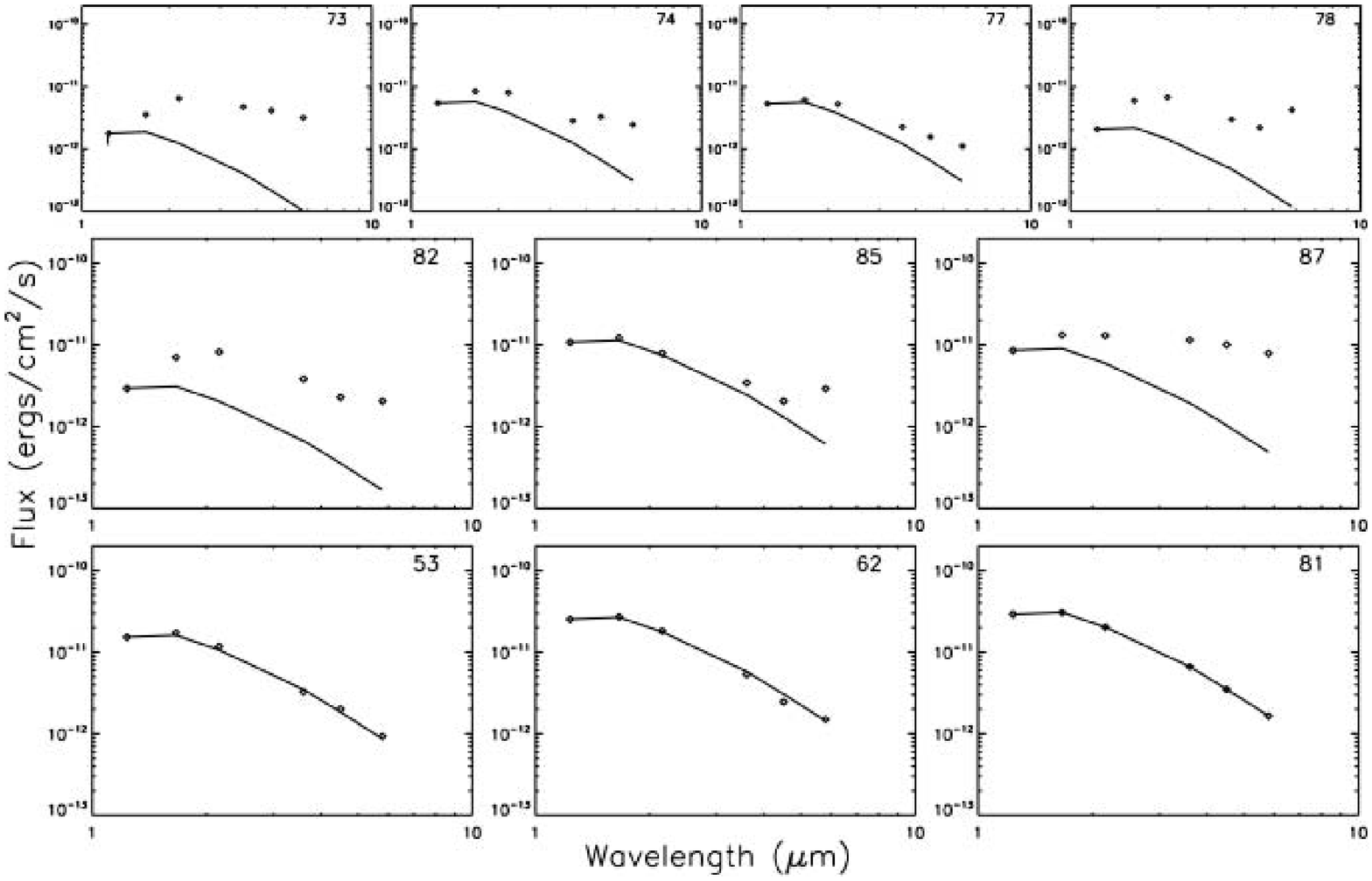}}

\clearpage

\clearpage

\clearpage

\clearpage

\clearpage

\clearpage

\end{document}